\documentclass[twocolumn]{IEEEtran}

\IEEEoverridecommandlockouts
\usepackage{cite}
\usepackage{amssymb,amsmath,mathtools}
\usepackage[utf8]{inputenc}
\usepackage{tikz}
\usepackage{cancel}
\usepackage{tabto}
\usepackage{bm}
\usepackage{float}
\usepackage{rotating} 
\usepackage{array}               
\usepackage{pdfpages}
\usepackage{pgfplots}
\usepackage{graphbox,graphicx}     
\usepackage{subfig}
\usepackage{enumitem}
\usepackage{floatflt}   
\usepackage{subfiles}
\usepackage{multirow}
\usepackage{colortbl}
\usepackage{algorithm}
\usepackage{algpseudocode}
\usepackage{notation}
\usepackage{pgf}
\usepackage{tikz}
\usepackage[utf8]{inputenc}
\usepackage{grffile}
\usepackage[export]{adjustbox}

\ifx\pgfplothandlerjumpmarkmid\undefined
  \def\pgfplothandlerjumpmarkmid{123}
\fi
\ifx\pgfplothandlerconstantlinetomarkmid\undefined
  \def\pgfplothandlerconstantlinetomarkmid{123}
\fi
\makeatletter
\ifx\pgf@remember@layerlist@globally\undefined
  \def\pgf@remember@layerlist@globally{123}
\fi
\ifx\pgf@restore@layerlist@from@global\undefined
  \def\pgf@restore@layerlist@from@global{123}
\fi

\usetikzlibrary{plotmarks}
\usepgfplotslibrary{polar}
\usetikzlibrary{arrows,automata}
\usetikzlibrary{positioning}
\usetikzlibrary{quotes,angles}
\pgfplotsset{compat=newest}
\usetikzlibrary{plotmarks}
\usetikzlibrary{arrows.meta}
\usetikzlibrary{spy}
\usepgfplotslibrary{patchplots}
\usetikzlibrary{calc,plotmarks,patterns,arrows.meta,fit,intersections,backgrounds}
\usepgfplotslibrary{colormaps,patchplots}
\usetikzlibrary{decorations.text}
\usetikzlibrary{decorations.pathmorphing}
\usetikzlibrary{spy}

\newcommand{\ist}{\hspace*{.3mm}}
\newcommand{\rmv}{\hspace*{-.7mm}}
\newcommand{\iist}{\hspace*{1mm}}
\newcommand{\rrmv}{\hspace*{-1mm}}
\newcommand{\rrrmv}{\hspace*{-3mm}}

\allowdisplaybreaks
\title{\huge Message Passing-Based 9-D Cooperative Localization and Navigation with Embedded Particle Flow}
\author{Lukas Wielandner\IEEEauthorrefmark{1}\IEEEauthorrefmark{2}, Erik Leitinger\IEEEauthorrefmark{1}, Florian Meyer\IEEEauthorrefmark{3}, Klaus Witrisal\IEEEauthorrefmark{1}\IEEEauthorrefmark{2}\\[2mm] 
	\IEEEauthorrefmark{1} Graz University of Technology \\ 
	\IEEEauthorrefmark{2} Christian Doppler Laboratory for Location-Aware Electronic Systems \\ 
	\IEEEauthorrefmark{3} University of California San Diego
	\thanks{This work was supported in part by the Christian Doppler Research Association; the
		Austrian Federal Ministry for Digital and Economic Affairs; and the National
		Foundation for Research, Technology, and Development.}
}

\begin{document}
\maketitle

%\mainmatter
\renewcommand{\baselinestretch}{0.99}\small\normalsize
%%%%%%%%%%%%%%%%%%%%%%%%%%%%%%%%%%%%%%%%%%%%
\begin{abstract}
	
Cooperative localization (CL) is an important technology for innovative services such as location-aware communication networks, modern convenience, and public safety. We consider wireless networks with mobile agents that aim to localize themselves by performing pairwise measurements amongst agents and exchanging their location information. Belief propagation (BP) is a state-of-the-art Bayesian method for CL. In CL, particle-based implementations of BP often are employed that can cope with non-linear measurement models and state dynamics. However, particle-based BP algorithms are known to suffer from particle degeneracy in large and dense networks of mobile agents with high-dimensional states.	

This paper derives the messages of BP for CL by means of particle flow, leading to the development of a distributed particle-based message-passing algorithm which avoids particle degeneracy. Our combined particle flow-based BP approach allows the calculation of highly accurate proposal distributions for agent states with a minimal number of particles. It outperforms conventional particle-based BP algorithms in terms of accuracy and runtime. Furthermore, we compare the proposed method to a centralized particle flow-based implementation, known as the exact Daum-Huang filter, and to sigma point BP in terms of position accuracy, runtime, and memory requirement versus the network size. We further contrast all methods to the theoretical performance limit provided by the posterior Cram\'er-Rao lower bound (PCRLB). Based on three different scenarios, we demonstrate the superiority of the proposed method.

\end{abstract}

%\begin{IEEEkeywords}
%Bayesian estimation, cooperative localization, particle flow, message passing.
%\end{IEEEkeywords}

%%%%%%%%%%%%%%%%%%%%%%%%%%%%%%%%%%%%%%%%%%%%
\section{Introduction}
%%%%%%%%%%%%%%%%%%%%%%%%%%%%%%%%%%%%%%%%%%%%
Location awareness is crucial for various applications, such as Internet-of-Things, autonomous navigation, or public safety \cite{da2014internet,WitrisalSPM2016,win2018IoT, DiTaranto2014SPM}. Cooperative localization (CL) methods aim to estimate the locations of agents in a wireless sensor network, where agents can communicate among their neighbors and exchange information about their position \cite{patwari2005locating, ShenTIT2010part2, ShenJSAC2012, WinProc2018TheoreticalFoundCoopLoc,KulmerTWC2018,win2018IoT}. This leads to an improvement of the positioning accuracy as well as an increasing localizability \cite{ping2021MaxFlow} while preventing the use of high-density anchor deployment as needed for non-CL \cite{patwari2005locating, alsindi2007cooperative,ShenTIT2010part2,MeiJ_WCOML2020,MeiJ_COML2021,WuJ_SPL2022,ZhangJ_SENSOR2022}. In fact, the anchor infrastructure can be fully avoided when using multipath channel information contained in radio-signals \cite{KulmerTWC2018}. Due to the increased localizability, CL is more robust than non-CL since more information in the network can be used. This increased robustness is especially useful for scenarios with very uninformative measurement models such as RSS based localization \cite{MeiJ_COML2021,ZhangJ_SENSOR2022,wielandner2021rss,wielandner2021jointEstimation}. CL algorithms are scalable and can be implemented in a distributed manner, which makes them particularly useful for large-scale networks \cite{wymeersch2009cooperative, Dardari2015IndoorTracking,meyer2015distributed_Loc}. 
A further crucial aspect of CL is to track high-dimensional agent states accurately. This paper proposes a new method for this purpose where different state-of-the-art algorithms fail as described in the following.

\begin{figure}[t]
\centering
\includegraphics[scale=1]{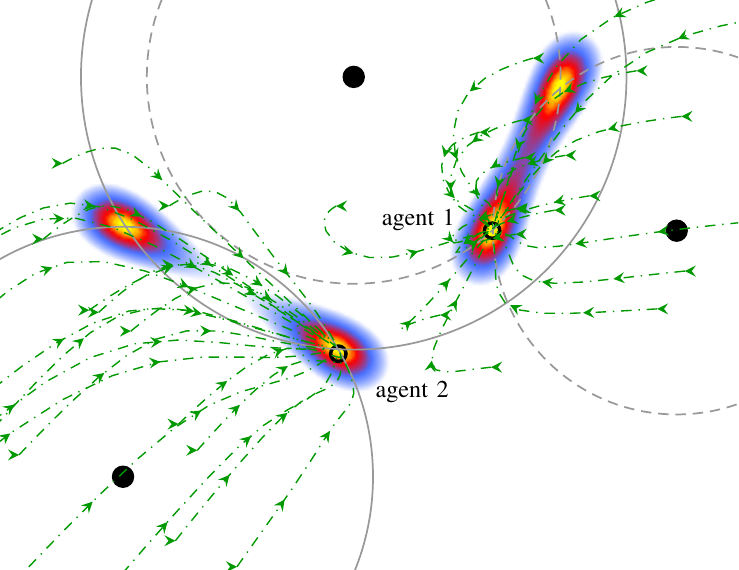}\\
\hfill
\includegraphics[scale=1]{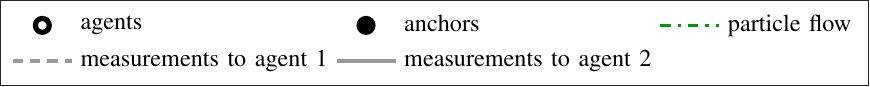}\hfill\\
\caption{Visualization of the particle flow (dash-dotted green lines) of two cooperating agents in the vicinity of three anchors. Each agent has only connections to two anchors (grey circles) indicated by the multimodal PDF of the agent positions (color map).
}
\label{fig:dyn_PF}
\end{figure}

\subsection{State-of-the-Art}

Promising methods for CL are based on the framework of factor graphs (FGs) and message-passing (MP) calculations, which can be categorized into mean-field message-passing-based methods \cite{cakmak2016cooperative,pedersen2011variational} and belief propagation (BP)-based methods \cite{ihler2005nonparametric, wymeersch2009cooperative,lien2012comparison, meyer2013sigma, meyer2015distributed_Loc, wielandner2021rss,meyer2021neuralEnhancedBP}. In particular, BP-based methods are known to provide accurate solutions to high-dimensional Bayesian estimation problems efficiently by executing message-passing on a cyclic FG. The sum-product rule is used to compute approximations ("beliefs") of the marginal posterior probability density functions (PDFs) of agent positions \cite{wymeersch2009cooperative, meyer2015distributed_Loc}. BP-based methods are very flexible and have been successfully applied to many diverse applications as for example radio signal-based simultaneous localization and mapping (SLAM) \cite{Leitinger2019BP_MP_SLAM, LeiGreWit:ICC2019, LeiMey:Asilomar2020_DataFusion}, multiobject tracking \cite{MeyKroWilLauHlaBraWinProIEEE2018,Gaglione2022FusionMOT,MeyWil:TSP2021}, and cooperative multiobject tracking \cite{BraGagSolRicGabLepNicWilBraWin:JSP2022}. Their excellent scalability and distributed nature make BP-based algorithms a powerful tool for CL on large-scale networks \cite{wymeersch2009cooperative, Dardari2015IndoorTracking, meyer2015distributed_Loc}. BP-based methods are categorized into parametric BP algorithms \cite{meyer2013sigma} and non-parametric BP algorithms \cite{ihler2005nonparametric}. Since the measurement models are usually non-linear and the calculations of the messages and beliefs cannot be evaluated in closed form, it is common to use non-parametric BP algorithms, resorting to conventional bootstrap particle-based implementations \cite{meyer2015distributed_Loc, wielandner2021rss}. A common drawback of such methods is the curse of dimensionality, a known problem of sample-based estimation in high dimensions, and the presence of informative measurements. The curse of dimensionality can lead to particle collapse, also known as particle degeneracy \cite{bengtsson2008curse}. It can often only be avoided by using an infeasible number of particles to represent the state accurately. Since the required memory and the computational demand are proportional to the number of particles, new strategies need to be developed for online estimation. A common approach to avoid particle degeneracy is to design an accurate proposal distribution or to make use of regularization \cite{musso2001improving,arulampalam2002tutorial,meyer2015distributed_Loc}. For the former, we have to address the problem of how to design accurate proposal distributions. Furthermore, regularization has to be treated very carefully since it can introduce biases if not correctly chosen. 

Recently, particle flow (PF) \cite{daum2010exact, daum2013nonzeroDiffusion, daum2017generalized, daum2018Gromov, Li2019FlowGMNoise, pmlr-v139-pal21b} was suggested for estimation in nonlinear systems with high-dimensional states and highly informative likelihood models. It is shown that the resulting PF particle filter is asymptotically optimal for nonlinear estimation problems and avoids particle degeneracy even for a relatively small number of particles. PF particle filters are successfully applied to multi-sensor localization \cite{li2017particle} and BP-based multi-target tracking \cite{ZhangMeyer2021MOT_PF} with the benefit that a significantly smaller number of particles are needed compared to bootstrap particle-based implementations. 
The main disadvantage of those methods is that they perform estimation based on the joint state. This increases the computational complexity excessively. Furthermore, some particle flow-based algorithms have an inherently large complexity, which provides an additional scaling by the number of used particles, for example, the localized EDH (LEDH) filter given in \cite{li2017particle} or the stochastic flow described in \cite{daum2018Gromov}. This makes it unattractive for large networks and does not allow for a distributed implementation.

\subsection{Contributions and Organization of the Paper}

This paper introduces a hybrid particle-based PF-BP message-passing algorithm for CL of mobile agents with 9-D states (three-dimensional position, velocity, and acceleration state vectors) and very informative measurement models. In this scenario, bootstrap particle-based BP methods that draw samples from predicted agent beliefs fail since an infeasible large number of particles is needed to represent the belief of agents accurately. Our approach avoids particle degeneracy using invertible PF \cite{li2017particle} to compute BP messages. Invertible PF enables the migration of particles towards regions of high probability, leading to an accurate approximation of BP messages with a relatively small number of particles. 
Therefore, the proposed algorithm combines the computational efficiency and scalability of BP methods with the benefits of the PF method. The proposed algorithm exploits the factorization structure of the cooperative localization problem. This leads to an inherent reduction of the number of dimensions per calculation, which also counteracts the particle degeneration problem and allows for a distributed implementation.  As an example, Figure~\ref{fig:dyn_PF} shows the particle flow of two cooperating agents, which are in the vicinity of three anchors. Since each agent has only connections to two anchors, the PDFs of the agent positions are multimodal. After considering the cooperative measurement, the particles flow to the ``correct mode'' of the posterior PDF, representing the ``true'' distribution of the agent positions. 

Numerical simulations demonstrate that the proposed PF-BP algorithm can significantly outperform a conventional bootstrap particle-based BP algorithm using sampling-importance-resampling (abbreviated with SIR-BP) \cite{meyer2015distributed_Loc}, a sigma point BP (SP-BP) algorithm \cite{meyer2013sigma}, and a particle-based exact Daum-Huang (EDH) filter (with a stacked state vector containing all agent state vectors) \cite{li2017particle} in terms of position accuracy. The results show that the proposed algorithm is Bayes-optimal in that it reaches the posterior Cram\'er-Rao lower bound (PCRLB)~\cite{patwari2005locating, TichavskyTSP1998}, which can also be expressed in the framework of the equivalent Fisher information matrix \cite{ShenTIT2010part2, ShenJSAC2012, WinProc2018TheoreticalFoundCoopLoc}. The proposed algorithm has much lower memory requirements than the SIR-BP algorithm since it needs a significantly smaller number of particles for the same level of position accuracy. The particle-based EDH filter calculates the matrix inversions and multiplications for the stacked state vector containing all agent states. Therefore, the memory requirements are also in favor of the proposed algorithm for the same number of particles. This is due to the fact that using PF-BP, the matrix inversions and multiplications reduce to the dimensions of a subset of the joint agent state. The key contributions of this paper can be summarized as follows.
\begin{itemize}
	\item We develop a distributed particle-based message-passing method for the CL of dynamic agents that computes BP messages using PF.
	\item We compare the proposed PF-BP method to state-of-the-art CL methods
	and demonstrate its superiority in terms of accuracy, runtime, and communication overhead.
	\item We demonstrate numerically that the proposed PF-BP method for CL can reach the PCRLB if the agents are localizable.
	\item We comprehensively analyze the investigated methods and highlight their benefits depending on different scenarios and applications.
\end{itemize}
In this work, we do not consider uncertainties beyond Gaussian noise, like missed detections, clutter/false alarm measurements, and data association uncertainty of measurements \cite{VenLeiTerWit:RadarCon2021, Gaglione2022FusionMOT, BraGagSolRicGabLepNicWilBraWin:JSP2022, meyer2020_scalabelDA}. This paper focuses on dynamic networks. The behavior of static networks can be analyzed
  by considering a single time step of the statistical model. This paper advances over the preliminary account of our method provided in the conference publication \cite{WielandnerICASSP2022} by (i) also considering the uncertainties of cooperating neighbor agents in the PF-BP belief update equations, (ii) a detailed description of the proposed algorithm, (iii) an extension to higher state dimensions, (iv) a comprehensive comparison to established state-of-the-art algorithms and to the theoretical performance limit in terms of the PCRLB.
The remainder of this paper is organized as follows. Section~\ref{sec:SystemModel} introduces the system and measurement model. We state the problem formulation in Section~\ref{sec:problem}. In Section~\ref{sec:reviewPF}, we provide a review of PF. In Section~\ref{sec:MP}, we describe the message-passing framework and explain the proposed method. The results of numerical experiments are reported in Section~\ref{sec:Results}. Section~\ref{sec:Conclusion} concludes the paper.

\subsubsection*{Notation}
Column vectors are denoted by boldface lowercase letters and matrices in boldface uppercase letters.
Random variables are indicated with sans serif, upright fonts and their realizations in serif, italic fonts as, for example, $\RV{x}$ and $\rv{x}$ and its respective realization as $\V{x}$ and $x$. We define the PDF of a continuous random variable as $f(\V{x})$. For a vector $\V{x}$, we indicate its transpose by $\V{x}\transp$ and the Euclidean norm by $\| \V{x} \|$. The mean value of a vector is denoted as $\overline{\V{x}}$. We will also use this notation to indicate the sample-based mean value and the minimum mean-square error (MMSE) estimate. The cardinality of a set $\Set{C}$ is defined as $| \Set{C} |$. Furthermore, we use the notation $\Set{C} \backslash \{i\}$ to indicate the exclusion of member $\{i\}$ from the set $\Set{C}$. The notation $\V{A} \otimes \V{B}$ denotes the Kronecker product between matrix $\V{A}$ and $\V{B}$, whereas $\odot$ indicates the Hadamard product. diag($\cdot$) stands for a diagonal matrix or a block diagonal matrix with elements on the main diagonal given by the elements or matrices in brackets, respectively. $\V{I}_m$ is an identity matrix of dimensions $m$. $[\V{X}]_{k:l,m:n}$ denotes a submatrix of $\V{X}$ containing $k$ to $l$ rows and $m$ to $n$ columns. The notation $[\V{x}]_{k:l}$ denotes a subvector of $\V{x}$ containing $k$ to $l$ elements. The time step $k$ is indicated by a superscript $^{(k)}$ whereas the $u$th message passing iteration with $^{[u]}$. $\nabla_{\V{x}}$ indicates the Nabla operator with respect to $\V{x}^{(k)}$.
\section{System Model}
\label{sec:SystemModel}

We consider a set of agents $\Set{C}$ and a set of anchors $\Set{A}$. The state of the agents is unknown, whereas the state of the anchors is exactly known. The number of agents and anchors is indicated by the cardinality of $\Set{C}$ and $\Set{A}$, respectively. We define two types of measurements: (i) measurements between agents and anchors $\rv{z}^{(k)}_{i,a}$ at time step $k$ with $i \in \Set{C}$ and $a \in \Set{A}^{(k)}_i$ where $\Set{A}^{(k)}_i \subseteq \Set{A}$ is the set of anchors that perform measurements to agent $i$ at time $k$ and (ii) measurements in-between agents $\rv{z}^{(k)}_{i,j}$ with $i \in \Set{C}$ and $j \in \Set{D}^{(k)}_i$ where $\Set{D}^{(k)}_i \subseteq \Set{C} \backslash \{i\}$ is the set of agents that cooperate with agent $i$ at time $k$. The stacked vector of all measurements for all time steps is written as $\RV{z} =[\rv{z}^{(1:K)}_{i,l}]_{i \in \Set{C}, l \in \Set{A}^{(1:K)}_i \cup \Set{D}^{(1:K)}_i}$ with $K$ being the total number of time steps. Each anchor has a fixed position which does not vary with time. The state of the $i$-th agent at time step $k$ is denoted as ${\RV{x}^{(k)}_i = [\RV{p}^{(k)\text{T}}_i \iist \RV{v}^{(k)\text{T}}_i \iist \RV{a}^{(k)\text{T}}_i]\transp} \in \mathbb{R}^{9\times1}$, where $\RV{p}^{(k)}_i \in \mathbb{R}^{3\times1}$, $\RV{v}^{(k)}_i \in \mathbb{R}^{3\times1}$, $\RV{a}^{(k)}_i \in \mathbb{R}^{3\times1}$ are, respectively, the position, velocity, and acceleration vectors. Thus, the number of dimensions per agent state is $N_{\text{D}}=9$.  We define the joint state of agent $i$ for all time steps as ${\RV{x}^{(1:K)}_i = [\RV{x}^{(1)\text{T}}_i \iist \dots \iist\RV{x}^{(K)\text{T}}_i]}\transp$. The states of the anchors are time-independent and assumed to be known. We write the state of the $a$-th anchor as ${\RV{x}_a = [\rv{p}_{x_a} \iist \rv{p}_{y_a} \iist \rv{p}_{z_a}]\transp}  \in \mathbb{R}^{3\times1}$. The vector $\RV{x}$ denotes the stacked vector of all agent and anchor states for all time steps. It is defined as ${\RV{x} = [\RV{x}^{(1:K)\text{T}}_1 \iist \dots \iist \RV{x}^{(1:K)\text{T}}_{|\Set{C}|}, \RV{x}\transp_{|\Set{C}|+1}\iist \dots \iist \RV{x}\transp_{|\Set{C}|+|\Set{A}|}]\transp}$. The $i$-th agent state $\V{x}^{(k)}_i$ is assumed to evolve according to a constant acceleration model given by
\begin{equation}
\V{x}^{(k)}_i = \V{F}\V{x}^{(k-1)}_i + \V{G}\V{u}^{(k-1)}
\label{eq:motionModel}
\end{equation}
with the state transition matrix $\V{F} \in \mathbb{R}^{9 \times 9}$ and the matrix $\V{G} \in \mathbb{R}^{9 \times 3}$ relating the state noise to the state variables. The state noise vector $\V{u}^{(k)} \in \mathbb{R}^{3 \times 1}$ is an independent and identically distributed (iid) sequence of 3-D Gaussian random vectors with standard deviation $\sigma_a$. The matrices are given as
\begin{equation}
\V{F} =\left[\begin{array}{ccc} 
1 & \Delta T & \frac{(\Delta T)^2}{2} \\ 
0 & 1  &\Delta T \\ 
0 & 0 & 1 \\ 
\end{array}\right] 
\otimes \V{I}_3
%\left[\begin{array}{ccc} 
%1 & 0 & 0 \\ 
%0 & 1  & 0\\ 
%0 & 0 & 1 \\ 
%\end{array}\right] 
\end{equation}
and
\begin{equation}
\V{G} =\left[\begin{array}{c} 
\frac{(\Delta T)^2}{2}  \\ 
\Delta T  \\ 
1 \\ 
\end{array}\right]
\otimes \V{I}_3\ist.
%\left[\begin{array}{ccc} 
%1 & 0 & 0 \\ 
%0 & 1  & 0\\ 
%0 & 0 & 1 \\ 
%\end{array}\right].
\end{equation}
Given the motion model, we can define the state transition probability and define the joint prior PDF for all agent states up to time $K$ using common statistical independence assumptions \cite{wymeersch2009cooperative,meyer2015distributed_Loc} as
\begin{equation}
f(\V{x}^{(1:K)}) = \prod_{k = 1}^K \prod_{i \in \mathcal{C}} f(\V{x}^{(0)}_i) f(\V{x}^{(k)}_i|\V{x}^{(k-1)}_i)\ist.
\label{eq:prior}
\end{equation}
The joint posterior PDF up to time $K$ is given as
\begin{align}
f(\V{x}^{(1:K)}|\V{z}^{(1:K)}) &\propto f(\V{z}^{(1:K)}|\V{x}^{(1:K)}) f(\V{x}^{(1:K)}).
\label{eq:contJointPost}
\end{align}
By assuming that measurements between nodes and time steps are independent of each other \cite{wymeersch2009cooperative,meyer2015distributed_Loc}, we can factorize the joint likelihood function as
\begin{align}
f(\V{z}^{(1:K)}|\V{x}^{(1:K)}) = & \prod_{k = 1}^K \prod_{i \in \mathcal{C}}  \prod_{a \in \Set{A}^{(k)}_i} f(z^{(k)}_{i,a}|\V{x}^{(k)}_{i},\V{x}_{a}) \nonumber \\ 
& \times \prod_{j \in \mathcal{D}^{(k)}_i} \rrmv \rrmv f(z^{(k)}_{i,j}|\V{x}^{(k)}_{i},\V{x}^{(k)}_{j}).
\label{eq:lhf}
\end{align}
The joint posterior PDF can now be written in terms of its factorization by plugging \eqref{eq:prior} and \eqref{eq:lhf} into \eqref{eq:contJointPost}, which results in
\begin{align}
&f(\V{x}^{(1:K)}|\V{z}^{(1:K)}) \nonumber \\ 
&\hspace*{7mm}\propto \ \prod_{k = 1}^K \prod_{i \in \mathcal{C}} f(\V{x}^{(0)}_i) f(\V{x}^{(k)}_i|\V{x}^{(k-1)}_i)  \nonumber \\
&\hspace*{9mm} \times \rrmv \rrmv \prod_{a \in \Set{A}^{(k)}_i} \rrmv \rrmv f(z^{(k)}_{i,a}|\V{x}^{(k)}_{i},\V{x}_{a}) \rrmv \rrmv \prod_{j \in \mathcal{D}^{(k)}_i} \rrmv \rrmv f(z^{(k)}_{i,j}|\V{x}^{(k)}_{i},\V{x}^{(k)}_{j}).
\label{eq:joint_post}
\end{align} 
We use distance measurements to infer the state of the agents. A measurement between two agents or between an agent and an anchor with indices $i$ and $j$, respectively, is given by 
\begin{align}
z^{(k)}_{i,j} = h(\V{x}^{(k)}_i,\V{x}^{(k)}_j) + n^{(k)}_{i,j}
\label{eq:measModel}
\end{align}
where $h(\V{x}^{(k)}_i,\V{x}^{(k)}_j) = \| \V{p}^{(k)}_j - \V{p}^{(k)}_i \|$. The measurement noise $\rv{n}_{i,j}$ is iid across $i$ and $j$, zero-mean, Gaussian with variance $\sigma^2$.

\section{Problem Formulation}
\label{sec:problem}

We aim to estimate mobile agent states $\V{x}^{(k)}_{i}$ cooperatively. Our Bayesian approach determines the marginal posterior PDF $f(\V{x}^{(k)}_i|\V{z}^{(1:k)})$ based on all measurements $\V{z}^{(1:k)}$ up to time $k$. Estimates of the agent state $\V{x}^{(k)}_{i}$ are obtained by the minimum mean-square error (MMSE) estimator \cite[Ch.~4]{Kay1993} given by
\begin{align}
	\overline{\V{x}}^{(k)}_{i} = \int \V{x}^{(k)}_i f(\V{x}^{(k)}_i|\V{z}^{(1:k)}) \text{d}\V{x}^{(k)}_i.
\end{align}
Since direct marginalization of the joint posterior in \eqref{eq:joint_post} typically cannot be evaluated in closed form, usually bootstrap particle-based BP \cite{KschischangTIT2001, Loeliger2004SPM} implementations are chosen to approximate the marginal PDFs. This conventional particle-based implementation suffers from particle degeneracy \cite{bengtsson2008curse} when agent states are high-dimensional, or measurements are very informative. Particle degeneracy leads to a ``wrong'' representation of agent beliefs that deteriorates the convergence behavior and performance of the particle-based BP algorithms. To overcome this issue, we propose a hybrid PF-BP algorithm. Before the proposed algorithm is introduced, a short review of the PF method is presented.

\section{Review of Particle Flow}
\label{sec:reviewPF}
In the case of a nonlinear measurement model as in \eqref{eq:measModel}, the posterior PDF $f(\V{x}|\V{z}) \propto f(\V{z}|\V{x})f(\V{x})$ is often approximated by a set of weighted samples $\{ w^{m},\V{x}^{m}\}^{M}_{m=1}$ with $\sum_{m=1}^M w^m \rmv=\rmv 1$ and the number of samples $M$. They are calculated based on the importance sampling principle \cite{arulampalam2002tutorial} as
\begin{equation}
w^m \propto \frac{f(\V{z}|\V{x}^m)f(\V{x}^m)}{q(\V{x}^m|\V{z})}
\end{equation}
with the proposal PDF $q(\V{x}|\V{z})$, from which the set of particles $\{ \V{x}^{m}\}^{M}_{m=1}$ is drawn. The only restriction to the proposal PDF is that it has to have the same support as the posterior PDF and heavier-tails \cite{DouFreGor:01}, i.e., it is less informative. Otherwise, it can be arbitrary. Importance sampling can provide an arbitrarily good approximation of the posterior PDF by choosing $M$ sufficiently large. Even though importance sampling is asymptotically optimal, if $q(\V{x}|\V{z})$ is correctly chosen, it is often infeasible to implement due to the large number of particles required for correct state estimation in high-dimensions.

\subsection{Derivation of the PF Equation}

Particle flow is an approach that migrates particles from the prior PDF to the posterior PDF by solving a partial differential equation \cite{daum2010exact, daum2013nonzeroDiffusion, li2017particle, daum2018Gromov, crouse2020consideration}. The particle flow is described by making use of the homotopy property and the Fokker-Planck equation (FPE) \cite{risken1996fokker}. The FPE is used to find a flow of particles that is equivalent to the flow of the probability density according to the log-homotopy function for the joint state $\RV{x}^{(k)}$ at time $k$. The log-homotopy function is given by \cite{daum2010exact,li2017particle}
\begin{align}
\text{log}f(\V{x}^{(k)};\lambda) &= \text{log}f(\V{x}^{(k)}|\V{x}^{(k-1)}) \nonumber \\[.5mm]
&\hspace{3mm} + \lambda \text{log} f(\V{z}^{(k)}|\V{x}^{(k)}) - \text{log} Z(\lambda)
\label{eq:log-homotopy}
\end{align}
where $\lambda \in [0,1]$ is the pseudo time of the flow process, $f(\V{x}^{(k)};\lambda)$ is the pseudo posterior during the flow process at time $\lambda$, and $Z(\lambda)$ is the evidence. We want to mention that $Z(\lambda = 0) = 1$.  The log-homotopy function describes a continuous and smooth deformation of the distribution starting from the prior PDF $f(\V{x}^{(k)}|\V{x}^{(k-1)})$, i.e., $\text{log}f(\V{x}^{(k)};0) = \text{log}f(\V{x}^{(k)}|\V{x}^{(k-1)})$ to finally result in the posterior PDF $\text{log}f(\V{x}^{(k)};1) \propto \text{log}f(\V{x}^{(k)}|\V{x}^{(k-1)}) + \text{log} f(\V{z}^{(k)}|\V{x}^{(k)})$.

It is assumed that the flow follows a stochastic differential equation of the form of \cite{daum2010exact,daum2013nonzeroDiffusion}
\begin{equation}
d\V{x}^{(k)} = \V{\zeta}(\V{x}^{(k)},\lambda)d\lambda + dw\ist .
\label{eq:stochPDE}
\end{equation} 
A detailed derivation of the flow equations can be found in Appendix~\ref{app:A}.

\subsection{Exact Daum-Huang (EDH) Filter}
\label{subsec:EDH}
This filter estimates the joint agent state $\RV{x}^{(k)}$ for each time step $k$. We review it since it will be a reference method and a fundamental cornerstone of our proposed approach. 

An analytic solution for $\V{\zeta}(\V{x}^{(k)},\lambda)$ in \eqref{eq:ODE}, given in Appendix~\ref{app:A}, can be found for Gaussian distributions \cite{daum2010exact}, resulting in the EDH filter \cite{daum2010exact,li2017particle}. To satisfy these conditions, we approximate the prior PDF as Gaussian distributed where $\V{R}^{(k)}$ and $\V{P}^{(k|k-1)}$  are the measurement noise covariance matrix and the predicted covariance matrix of the joint state at time $k$, respectively. The solution for $\V{\zeta}(\V{x}^{(k)},\lambda)$, according to the EDH filter, is given by \cite{crouse2020consideration}
\begin{equation}
\V{\zeta}(\V{x}^{(k)},\lambda) = \V{A}^{(k)}_\lambda \V{x}^{(k)} + \V{c}^{(k)}_\lambda\ist.
\label{eq:EDH}
\end{equation}
A detailed description of the EHD filter and its implementation can be found in Appendix~\ref{app:B}, providing also the solution for \eqref{eq:stochPDE}. We would like to point out that the EDH in this form can only be implemented in a centralized manner.

\section{Message Passing Algorithms and Proposed Method}
\label{sec:MP}
In a Bayesian framework, we estimate the position of each agent based on the marginal posterior PDFs. Since a direct marginalization of the joint posterior \eqref{eq:joint_post} is often infeasible, we perform message passing (MP) by means of the sum-product-algorithm rules on the factor graph that represents our statistical model. This so-called ``belief propagation (BP)'' yields approximations (``beliefs'') of the marginal posterior PDFs in an efficient way \cite{KschischangTIT2001, Loeliger2004SPM}. It gives the exact marginal PDFs for a tree-like graph but provides only an approximate marginalization if the underlying factor graph has cycles \cite{KschischangTIT2001}. In this case, the BP message passing becomes iterative, and there exist different orders in which the messages can be calculated. We have chosen that in each iteration, the beliefs of all agents $i\in \Set{C}$ are updated in parallel. In the following section, we derive the MP scheme based on the factor graph in Figure~\ref{fig:FG}. In Section~\ref{sec:BP}, we shortly present the standard particle-based implementation of BP, whereas in Section~\ref{sec:BPPF}, we state the proposed method based on the same MP scheme.

\subsection{BP Message Passing}
\label{subsec:MP}
\begin{figure}[t]
\includegraphics[scale=0.9]{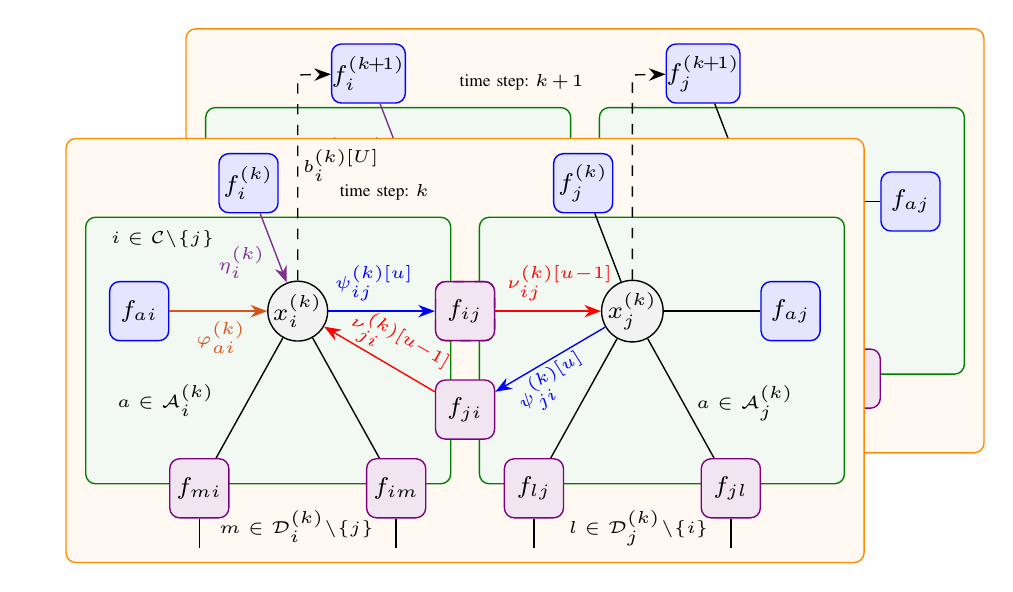}\\
\hfill
\caption{This figure shows a graphical representation of the system model in terms of a factor graph at time step $k$. The notation $\Set{D}^{(k)}_m \backslash \{l\}$ means all members of  $\Set{D}^{(k)}_m$ except $l$. We use the short hand notation: $b^{(k)[U]}_{i} \triangleq b^{[U]}(\V{x}^{(k)}_i)$, $\eta^{(k)}_{i} \triangleq \eta(\V{x}^{(k)}_{i})$, $\nu^{(k)[u-1]}_{ji} \triangleq \nu^{[u-1]}_{j}(\V{x}^{(k)}_{i})$, $\varphi^{(k)}_{ai} \triangleq \varphi_{a}(\V{x}^{(k)}_{i})$ and $\psi^{(k)[u]}_{ij} \triangleq \psi^{[u]}_{i}(\V{x}^{(k)}_{j})$. Factors $f_{ij}$ change depending on a distributed or centralized processing scheme.}
\vspace{-3mm}
\label{fig:FG}
\end{figure}

Based on the factor graph in Figure~\ref{fig:FG}, we define the MP scheme to approximate the marginal posterior PDFs. For a better readability, we use the following shorthand notation: In a \textit{distributed} implementation of BP, the factor $f_{ij}~\triangleq~f(z^{(k)}_{i,j}|\V{x}^{(k)}_{i},\V{x}^{(k)}_{j})$ represents the likelihood function with respect to the involved agents $i$ and $j$ at time $k$ since only measurement $z^{(k)}_{i,j}$ is available at node $\V{x}^{(k)}_{i}$. Therefore $f_{ij} \neq f_{ji}$. In a \textit{centralized} implementation, both measurements between agent $i$ and $j$ at time $k$ are available. Therefore the factor is given as the product of the likelihood of both measurements as $f_{ij}~\triangleq~f(z^{(k)}_{i,j}|\V{x}^{(k)}_{i},\V{x}^{(k)}_{j}) f(z^{(k)}_{j,i}|\V{x}^{(k)}_{i},\V{x}^{(k)}_{j})$, which results in $f_{ij} = f_{ji}$. The factor $f^{(k)}_{i}~\triangleq~f(\V{x}^{(k)}_i|\V{x}^{(k-1)}_i)$ corresponds to the state transition PDF. At time $k=0$ it corresponds to the prior PDF $f(\V{x}^{(0)}_i)$. The factor $f_{ai}~\triangleq~ f(z_{i,a}|\V{x}^{(k)}_{i},\V{x}_{a})$ represents information from an anchor measurement. 
Since the factor graph has loops, we use an iterative MP scheme to approximate the marginal PDF (belief) of agent state $i$ at time step $k$. We define the belief at MP iteration $u~\in~\{1,\dots, U\}$ as the product of all incoming messages as
\begin{equation}
b^{[u]}(\V{x}^{(k)}_{i}) = \eta(\V{x}^{(k)}_{i}) \prod_{a \in \Set{A}^{(k)}_i} \varphi_{a}(\V{x}^{(k)}_{i}) \prod_{j \in \Set{D}^{(k)}_i} \nu^{[u-1]}_{j}(\V{x}^{(k)}_{i}).
\label{eq:BP_messagepassing1}
\end{equation}
The messages are defined in the following manner: 
The message representing the state transition of agent $i$ is given as
\begin{equation}
\eta(\V{x}^{(k)}_{i}) = \int f(\V{x}^{(k)}_i|\V{x}^{(k-1)}_i) b^{[U]}(\V{x}^{(k-1)}_{i}) d\V{x}^{(k-1)}_{i}
\end{equation}
whereas the message from anchor $a$ to agent $i$ is
\begin{align}
\varphi_{a}(\V{x}^{(k)}_{i}) & = \int f(z_{i,a}|\V{x}^{(k)}_{i},\V{x}_{a}) \delta(\V{x}_{a} - \V{x}_{\text{true},a}) d\V{x}_{a} \nonumber \\
& =  f(z_{i,a}|\V{x}^{(k)}_{i};\V{x}_{\text{true},a})
\end{align}
where $\V{x}_{\text{true},a}$ corresponds to the true position of anchor $a$.
Using the extrinsic information $\psi_i^{[u-1]}(\V{x}^{(k)}_{j})$ from the cooperative agent $j$, the messages of the cooperative part can be written in the form of
\begin{equation}
\nu^{[u-1]}_{j}(\V{x}^{(k)}_{i}) = \int f(z^{(k)}_{i,j}|\V{x}^{(k)}_{i},\V{x}^{(k)}_{j}) \psi_i^{[u-1]}(\V{x}^{(k)}_{j}) d\V{x}^{(k)}_{j} 
\end{equation}
for a \textit{distributed} implementation since only measurement $z^{(k)}_{i,j}$ is available at node $\V{x}^{(k)}_{i}$. In a \textit{centralized} manner, it is given as
\begin{align}
\nu^{[u-1]}_{j}(\V{x}^{(k)}_{i}) & = \int f(z^{(k)}_{i,j}|\V{x}^{(k)}_{i},\V{x}^{(k)}_{j}) \nonumber \\ 
& \times f(z^{(k)}_{j,i}|\V{x}^{(k)}_{i},\V{x}^{(k)}_{j}) \psi_i^{[u-1]}(\V{x}^{(k)}_{j}) d\V{x}^{(k)}_{j} 
\end{align}
since both measurements between agent $i$ and $j$ at time $k$ are available.
The extrinsic information is given as
\begin{equation}
\psi_i^{[u]}(\V{x}^{(k)}_{j}) = \eta(\V{x}^{(k)}_{j}) \prod_{a \in \Set{A}^{(k)}_j} \varphi_{a}(\V{x}^{(k)}_{j}) \prod_{l \in \Set{D}^{(k)}_j \backslash \{i\}} \nu^{[u-1]}_{l}(\V{x}^{(k)}_{j})
\end{equation}
where the notation $\Set{D}^{(k)}_j \backslash \{i\}$ indicates that $i$ is excluded from the set $\Set{D}^{(k)}_j$. It is very common to approximate the extrinsic information by the corresponding belief, resulting in $\psi_i^{[u]}(\V{x}^{(k)}_{j}) \approx b^{[u]}(\V{x}^{(k)}_{j})$ \cite{wymeersch2009cooperative,lien2012comparison,meyer2015distributed_Loc}. This reduces the computational complexity significantly since it avoids calculating the extrinsic information, which is different for each cooperating agent pair. An additional benefit is that it also reduces the communication between the agents since exchanging extrinsic information requires point-to-point communication, whereas the belief can be broadcast \cite{wymeersch2009cooperative,lien2012comparison,meyer2015distributed_Loc}. Throughout the paper, we use the approximation of extrinsic information.

The agent marginal PDF $f(\V{x}^{(0:k)}_i|\V{z}^{(1:k)})$ is approximated up to a normalization constant by the belief $b^{[u]}(\V{x}^{(k)}_i)$. We estimate the state of the $i$-th agent at the end of the MP iterations according to the MMSE estimator \cite{Kay1993} as
\begin{equation}
\bar{\V{x}}^{(k)}_i= \int \V{x}_i b^{[U]}(\V{x}^{(k)}_i) \text{d}\V{x}^{(k)}_i.
\label{eq:mmse}
\end{equation}

\subsection{SIR-BP Algorithm}
\label{sec:BP}
We represent the belief at MP iteration $u$ with a weighted set of particles $\{w_i^{(k)[u],m}, \V{x}_i^{(k)[u],m}\}_{m=1}^{M}$. For further insights, please refer to \cite{meyer2015distributed_Loc}.
After each iteration $u$, we use systematic resampling \cite{arulampalam2002tutorial} to approximate the belief of the $i$th agent state by a set of equally weighted particles as $\{1/M, \V{x}_i^{(k)[u],m}\}_{m=1}^{M}$, where $M$ is the number of particles. To avoid particle degeneracy after resampling, we can use regularization to convolve the resampled set of particles with a kernel that could be estimated or predefined \cite{MeyerJSAC2015}. I.e., the $m$-th particle $\acute{\V{x}}_i^{(k)[u],m}$ is drawn from a Gaussian distribution with a mean value of $\V{x}_i^{(k)[u],m}$ and a covariance of $\Sigma_r$.

\subsection{PF-BP Algorithm}
\label{sec:BPPF}
This approach uses the same BP MP to approximate the marginal PDF of the state as mentioned in Section~\ref{subsec:MP}. The only difference is that instead of a point-wise multiplication of the incoming messages at a variable node, we use particle flow to determine the product of the messages. We represent the agent state $i$ at time $k$ by a set of equally weighted particles $\{1/M, \V{x}_i^{(k),m}\}_{m=1}^{M}$. In the following, we present the particle-based implementation of PF-BP.

Comparing to Section~\ref{subsec:EDH} and Appendix~\ref{app:B}, the flow of the $m$-th particle, representing the approximate marginal posterior PDF of agent $i$ at time step $k$, pseudo-time step $\lambda_l$ and message passing iteration $u$ is given as
\begin{equation}
\V{x}_{\lambda_l,i}^{(k)[u],m} = \V{x}_{\lambda_{l-1},i}^{(k)[0],m} + \tilde{\V{\zeta}}(\V{x}_{\lambda_{l-1},i}^{(k)[0],m},\V{{x}}_{\rightarrow i}^{(k)[u-1],m},\lambda_l) \Delta_l.   \label{eq:int_flow_BP}
\end{equation}
This recursive equation represents the particle-based multiplication of the incoming messages $\varphi_{a}(\V{x}^{(k)}_{i})$ and $\nu^{[u-1]}_{j}(\V{x}^{(k)}_{i})$ for $a \in \Set{A}^{(k)}_i$ and $j \in \Set{D}^{(k)}_i$. The message $\eta(\V{x}^{(k)}_{i})$ is obtained by propagating the particle representation through the motion model. Therefore, we define the $m$-th particle, drawn from the proposal PDF as $\V{x}_{\lambda_l = 0,i}^{(k)[u=0],m} = \V{x}_{i}^{(k|k-1),m}$, being equal to the predicted particle by the motion model.

The variable $\V{{x}}_{\rightarrow i}^{(k)[u]}$ can be seen as a joint state representing the beliefs of agents that perform measurements to agent $i$ at time $k$, evaluated at MP iteration $u$. $\V{x}_{\rightarrow i}^{(k)[u],m}$ indicates the $m$-th particle of the stacked representation of this joint state. It will be explained in what follows. 
The particles represented in \eqref{eq:int_flow_BP} at $\lambda = 1$ do not exactly match the particles drawn from the corresponding proposal density. Therefore, we have to use the invertible flow, as mentioned in \cite{li2017particle} and recalculate the weights of the particles. This is done based on the particle representation at the end ($\lambda = 1$) and the beginning ($\lambda = 0$) of the flow as
\begin{align}
w_i^{(k)[u],m} \propto &  \frac{f(\V{x}_{\lambda = 1,i}^{(k)[u],m}|\V{x}_{i}^{(k-1),m})}{f(\V{x}_{\lambda = 0,i}^{(k)[u=0],m}|\V{x}_{i}^{(k-1),m})} \nonumber \\
& \times \rrrmv \rrrmv \displaystyle \prod_{j \in \Set{A}^{(k)}_i \cup \Set{D}^{(k)}_i } \rrrmv \rrrmv f(z_{i,j}|\V{x}_{\lambda = 1,i}^{(k)[u],m},\V{x}_{j}^{(k)[u-1],m}).
\label{eq:margWeights}
\end{align}
The belief of agent state $i$ at time $k$ and MP iteration $u\vspace*{1mm}$, given in \eqref{eq:BP_messagepassing1}, is represented by the weighted set of particles $\{w_i^{(k)[u],m}, \V{x}^{(k)[u],m}_{\lambda = 1,i}\}_{m=1}^{M}$. Using the weighted particle representation, we perform systematic resampling to approximate $b^{[u]}(\V{x}^{(k)}_i) \vspace*{0.3mm}$ by a set of particles with uniform weights $\{1/M, \V{x}_i^{(k)[u],m}\}_{m=1}^{M}$ where we again drop the index $\lambda$ to indicate the resampled particles. At this point, we want to mention that the final approximation of the marginal posterior PDF at MP iteration $U$ is indicated by $\{1/M, \V{x}_i^{(k),m}\}_{m=1}^{M}$, neglecting the MP index.

We introduce a new variable $\V{\chi}^{(k)[u]}_i$ that corresponds to the resampled set of particles. The covariance matrix of the belief of agent $i$ is indicated as $\V{P}_{i}^{(k)[u]}$. Even though it is possible, we do not determine $\V{P}_{i}^{(k)[u]}$ using the particle representation but based on the UKF update step as described in what follows. We chose this approach since it was observed that the particle representation could collapse after resampling. 

For each MP iteration $u$, we let the particles of the agent state $i$ flow for all $\lambda \vspace*{0.4mm}$-steps.  In addition, we define $\V{{x}}_{\rightarrow i}^{(k)[u-1]} = [\V{\chi}^{(k)[u-1]}_j]_{j \in \Set{D}^{(k)}_i}$ , which indicate the states of agents that perform a measurement to agent $i$ at time $k$, and the sample-based mean value of it as $\overline{\V{x}}_{\rightarrow i}^{(k)[u-1]}$. The states of the cooperating agents are represented by their beliefs at the previous iteration $[u-1]$. Furthermore we define the stacked representation of the joint state of agent $i$ at pseudo time step $\lambda_{l-1} \vspace*{0.4mm}$ and its cooperative partners at MP iteration $u$ as $\V{\beta}^{(k)[u]}_{\lambda_{l-1},i} = [\V{x}_{\lambda_{l-1},i}^{(k)[0] \text{T}},\V{{x}}_{\rightarrow i}^{(k)[u-1] \text{T}}]\transp$ and its sample-based mean value as $\overline{\V{\beta}}^{(k)[u]}_{\lambda_{l-1},i} = [\overline{\V{x}}_{\lambda_{l-1},i}^{(k)[0] \text{T}},\overline{\V{x}}_{\rightarrow i}^{(k)[u-1] \text{T}}]\transp$. With that, we can write the drift of each particle $m$ as
\begin{equation}
\V{\zeta}(\V{x}_{\lambda_{l-1},i}^{(k)[0],m},\V{{x}}_{\rightarrow i}^{(k)[u-1],m},\lambda_l) = \V{A}_i \V{\beta}^{(k)[u],m}_{\lambda_{l-1},i} + \V{c}_i
\label{eq:EDH_gradient} 
\end{equation}
with $\V{A}_i \triangleq \V{A}(\V{x}_{\lambda_{l-1},i}^{(k)[0]},\V{{x}}_{\rightarrow i}^{(k)[u-1]},\lambda_l)$ and $\V{c}_i \triangleq \V{c}(\V{x}_{\lambda_{l-1},i}^{(k)[0]},\V{\hat{x}}_{\rightarrow i}^{(k)[u-1]},\lambda_l)$. For the flow update in \eqref{eq:int_flow_BP}, $\tilde{\V{\zeta}}(\cdot)$ consists of the first $N_{\text{D}}$ elements of $\V{\zeta}(\cdot)$ in \eqref{eq:EDH_gradient}. This corresponds to the drift of the marginal distribution of agent state $i$, since the dimension of $\V{x}^{(k)}_i$ is $N_{\text{D}}$. The flow of the mean value of the agent state is similar to \eqref{eq:int_flow_BP} where we replace the particle representation of the agent state with the mean values as in \eqref{eq:int_flow_mean}.

With that in mind, we can define $\V{A}_i$ and $\V{c}_i$ as
\begin{align}
\V{A}_i = & -\frac{1}{2} \tilde{\V{P}}_i \V{H}^{(k) \text{T}}_i (\lambda_l \V{H}^{(k)}_i \tilde{\V{P}}_i \V{H}^{(k) \text{T}}_i + \V{R}^{(k)}_i)^{-1} \V{H}^{(k)}_i \label{eq:A_i} \\
\V{c}_i = & (\V{I}_{N_{\text{D}}(|\Set{D}^{(k)}_i|+1)} \rmv + \rmv 2\lambda_l \V{A}_i) \left[(\V{I}_{N_{\text{D}}(|\Set{D}^{(k)}_i|+1)} \rmv + \rmv \lambda_l \V{A}_i) \right. \nonumber \\
& \times \left. \tilde{\V{P}}_i \V{H}^{(k) \text{T}}_i (\V{R}^{(k)}_i)^{-1}  (\V{z}_i \rmv - \rmv \V{\nu}_i) + \V{A}_i \overline{\V{\beta}}_{\lambda = 0,i}^{(k)[u]} \right] \label{eq:b_i}
\end{align}
with 
\begin{equation}
\V{\nu}_i = [h(\overline{\V{x}}_{\lambda_{l-1},i}^{(k)[0]},\overline{\V{{\vartheta}}}_{q}^{(k)})]_{q \in \Set{A}^{(k)}_i \cup \Set{D}^{(k)}_i } - \V{H}^{(k)}_i \overline{\V{\beta}}^{(k)[u]}_{\lambda_{l-1},i}
\end{equation}
where $\V{\nu}_i$ corresponds to the model mismatch due to the linearization and $ \vspace*{0.6mm}\overline{\V{\vartheta}}^{(k)} = [\V{x}_{\text{true}\transp,\Set{A}_i(1)},\dots,\V{x}_{\text{true},\Set{A}_i(|\Set{A}_i|)}\transp,\overline{\V{x}}_{\rightarrow i}^{(k)[u-1] \text{T}}]\transp \vspace*{0.6mm}$, $\V{z}_i = [z_{i,j}]_{j \in \Set{A}^{(k)}_i \cup \Set{D}^{(k)}_i}$, and $\overline{\V{x}}_{\lambda_{l},i}^{(k)[u]} = (1/M)\sum_{m=1}^M \V{x}_{\lambda_{l},i}^{(k)[u],m}$. In what follows, we define all other involved vectors and matrices.

The observation matrix $\V{H}^{(k)}_i$ has the dimensions $(|\Set{A}^{(k)}_i| + |\Set{D}^{(k)}_i|) \times N_{\text{D}}(1+|\Set{D}^{(k)}_i|)$, which is equivalent to the number of measurements of agent $i$ times the sum of the dimensions of all involved states. $\V{H}^{(k)}_i$ consists of the $N_{\text{D}}$-dimensional elements 
\begin{equation}
[\V{H}^{(k)}_i]_{\tilde{o},N_{\text{D}}\tilde{p}-N_{\text{D}}+1:N_{\text{D}} \tilde{p}} = \frac{\partial h(\V{x}^{(k)}_p,\V{x}^{(k)}_o)}{\partial \V{x}^{(k)}_p} \left|_{\V{x}^{(k)}_p = \hat{\overline{\V{\beta}}}_{\tilde{p}}} \right. \label{eq:H_j}
\end{equation}
for $p \in \{i\} \cup \Set{D}_i$, which is a sorted set with index $\tilde{p}$, representing the index of the cooperative partner, and the sorted set $o \in \Set{A}^{(k)}_i \cup \Set{D}^{(k)}_i$, with index $\tilde{o}$, determining the index of the $o$-th \vspace*{0.6mm} measurement. The derivative is evaluated at $\hat{\overline{\V{\beta}}}_{\tilde{p}} = [\overline{\V{\beta}}^{(k)[u]}_{\lambda_{l-1},i}]_{N_{\text{D}}\tilde{p}-N_{\text{D}}+1:N_{\text{D}} \tilde{p}}$.

The first three elements in \eqref{eq:H_j} correspond to the derivative with respect to the position coordinates. The following three elements correspond to the derivative with respect to the velocity coordinates, and the last three elements correspond to the derivative with respect to the acceleration coordinates. The elements containing the derivative with respect to velocity and acceleration are zero since the observation model depends only on the position.

\begin{algorithm}[t]
\caption{Proposed PF-BP Algorithm}\label{alg:PFBP}
\begin{algorithmic}[1]
\For{$i=1:|\Set{C}|$}
\State initialize Gaussian prior distribution with mean value \newline \hspace*{1.15em} 
$\overline{\V{x}}^{(0)}_i$ and covariance matrix $\V{P}^{(0)}_i$.
\State draw particles $\{1/M,\V{x}^{(0),m}_i\}^{M}_{m = 1}$ from prior \newline \hspace*{1.15em}  distribution
\EndFor
\For{k =1:K} 
\For{$i=1:|\Set{C}|$}
\State predict particles and covariance matrix according  \newline \hspace*{2.7em} to \eqref{eq:motionModel} and \eqref{eq:PFBP_prediction}.
\State determine sample-based mean value $\overline{\V{x}}^{(k)[0]}_{\lambda=0,i}$
\EndFor
\For{$u =1:U$}
\For{$i=1:|\Set{C}|$}
\State calculate flow according to \eqref{eq:int_flow_BP}  \newline \hspace*{4.1em} (using \eqref{eq:EDH_gradient}--\eqref{eq:P_i}) for all $\lambda$-steps
\State resample particles according to \eqref{eq:margWeights} to get \newline \hspace*{4.1em} $\{1/M, \V{x}_i^{(k)[u],m}\}_{m=1}^{M}$
\State determine sample-based mean value $\overline{\V{x}}^{(k)[u]}_{i}$
\State calculate $\V{P}^{(k)[u]}_i$ according to \eqref{eq:PFBP_update} at $\overline{\V{x}}^{(k)[u]}_{i}$
\State optional: regularization of resampled particles \newline \hspace*{4.1em} and $\V{P}^{(k)[u]}_{i}$ according to \eqref{eq:PFBP_reg}
\EndFor
\EndFor
\For{$i=1:|\Set{C}|$}
\State determine MMSE estimate according to \newline \hspace*{2.7em} sample-based mean value $\overline{\V{x}}^{(k)[U]}_{i}$
\EndFor
\EndFor
\end{algorithmic}
\end{algorithm}

The measurement noise covariance matrix $\V{R}^{(k)}_i$ has the dimensions $(|\Set{A}^{(k)}_i| + |\Set{D}^{(k)}_i|) \times (|\Set{A}^{(k)}_i| + |\Set{D}^{(k)}_i|)$ with $\sigma^2$ at the main diagonal and zeros elsewhere. We also define the block-diagonal covariance matrix of the involved states at time $k$ as
\begin{equation}
\tilde{\V{P}}_i = \text{diag}\left(\V{P}_{i}^{(k|k-1)},\dots,\V{P}_{m}^{(k)[u-1]},\dots\right)
\label{eq:P_i}
\end{equation} 
where $\V{P}_{i}^{(k|k-1)}$ is the predicted covariance matrix of agent state $i$ and $\V{P}_{m}^{(k)[u-1]}$ are the covariance matrices of the states of all other connected agents $m \in \Set{D}^{(k)}_i$ determined at flow time $\lambda=1$ of the previous MP iteration $u-1$. Similarly to \cite{li2017particle}, these covariance matrices are calculated, respectively, using a UKF covariance matrix prediction and update, i.e., 
\begin{align}
\V{P}_{i}^{(k|k-1)} & = \V{F}\V{P}_{i}^{(k-1)[U]}  \V{F}\transp + \V{Q} \label{eq:PFBP_prediction} \\
\V{P}_{i}^{(k)[u]}      & = \V{P}_{i}^{(k|k-1)}  - \tilde{\V{K}}^{[u]} \tilde{\V{P}}_{zz} \tilde{\V{K}}^{[u] \text{T}}\label{eq:PFBP_update}
\end{align}
where $\tilde{\V{K}}^{[u]}$ again represents the Kalman gain at MP iteration $u$ since it depends on the beliefs of the involved agent states, and $\tilde{\V{P}}_{zz}$ is the measurement covariance matrix. As discussed above, we perform systematic resampling at the end of each MP iteration resulting in $\{1/M, \V{x}_{i}^{(k)[u],m} \}_{m=1}^M$. Note that the covariance matrices $\V{P}_{i}^{(k)[u]}$ are calculated at sample-based mean value $\overline{\V{x}}_{i}^{(k)[u]}$. 
 In addition to the particles, we represent the marginal posterior PDF of agent $i$ at time $k$ and MP iteration $u$, with a mean value and a covariance matrix. At MP iteration $U$, we determine the MMSE estimate of each agent state according to the sample-based mean value of each agent state.
We use an exponentially spaced $\lambda$ as suggested in \cite{daum2013nonzeroDiffusion}, which results in a more accurate position estimate in our simulations compared to a linear spacing with the same number of steps. 
A summary of the particle-based implementation of PF-BP is provided in Algorithm~\ref{alg:PFBP}.

\begin{figure}[t]
\centering
\includegraphics[scale=1]{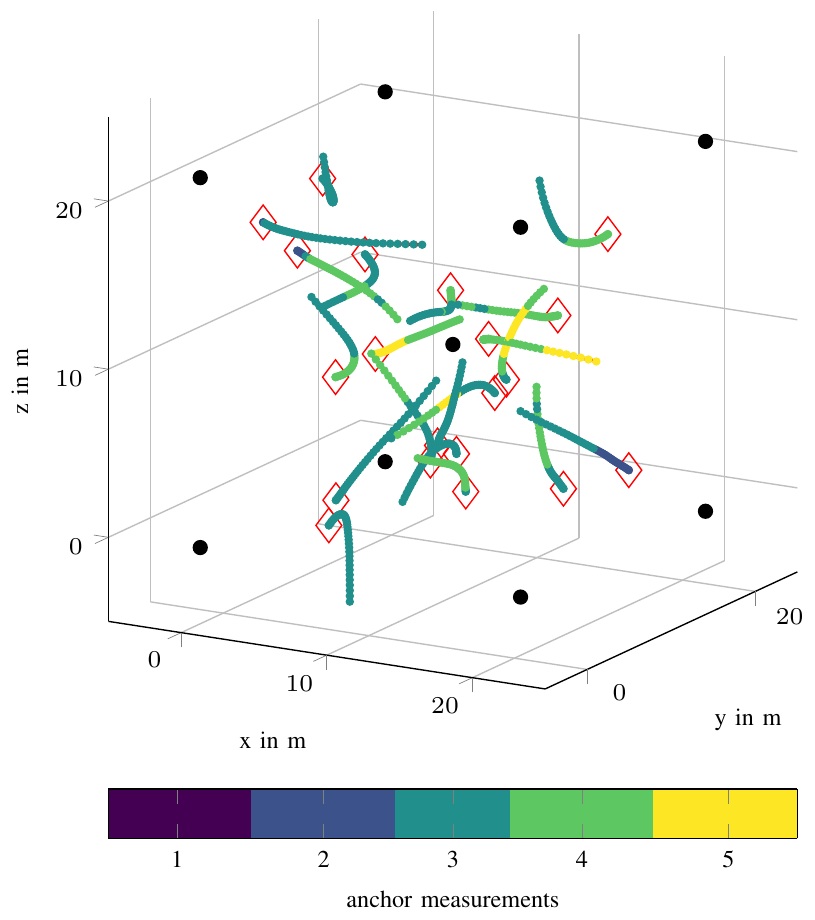}\\
\hfill
\caption{A realization of the trajectories for 20 agents. Anchors are given in black. The initial positions of the agents are marked with red diamonds, and the trajectory is given in red. The colored scatter points indicate how many connections an agent has to anchors along its trajectory. The communication range is $r_{\text{max}}=18$~m. Agents have at least one connection to an anchor at every time step.}
\label{fig:dyn_scen}
\end{figure} 

\section{Evaluation of Algorithms}
\label{sec:Results}
In this section, we evaluate the proposed algorithm based on dynamic networks for various network sizes and connectivities. We use a constant acceleration motion model in 9D (three-dimensional position, velocity, and acceleration state vectors) given in \eqref{eq:motionModel}. We compare the performance to a bootstrap particle-based BP algorithm (termed SIR-BP) described in Section~\ref{sec:BP}, a SP-BP algorithm \cite{meyer2013sigma}, and to a fully joint particle-based EDH filter \cite{li2017particle}. Furthermore, we show the theoretical performance limit w.r.t. the PCRLB \cite{patwari2005locating,TichavskyTSP1998}. We determine the performance in terms of the root-mean-square error (RMSE) of the MMSE estimates of position ($\text{RMSE}_\text{p}$), velocity ($\text{RMSE}_\text{v}$) and acceleration ($\text{RMSE}_\text{a}$), the cumulative frequency (CF) of the position error, and the runtime per time step. In addition, we show the probability of outage of the position error versus a position error threshold. The outage is defined as position errors above the position error threshold. The uncertainty of the measurement model is $\sigma = 0.1$~m. In the following simulations, we use 9 anchors and two different numbers of agents defined as $N_\text{agent} \in \{5,20\}$. The true agent positions are uniformly drawn for each realization in a volume of 20$\,$m $\times$ 20$\,$m $\times$ 20$\,$m. The true velocity of each agent is initialized with a unit vector in the direction of the center of the scenario, while the true acceleration is initialized with zero. The agent trajectories are generated in 3D based on a constant acceleration model given in \eqref{eq:motionModel} with $\Delta T = 0.1$~s and the standard deviation of $\V{u}^{(k)}$ is $\sigma_a$ = 0.15~m/s$^2$. The prior distribution for position (except for the SIR-BP algorithm), velocity and acceleration of each agent state $\RV{x}_i$ is initialized with a Gaussian distribution with a mean value of $\overline{\V{x}}^{(0)}_i = [\overline{\V{p}}^{(0)\text{T}}_i \overline{\V{v}}^{(0)\text{T}}_i \overline{\V{a}}^{(0)\text{T}}_i]\transp$, which will be defined later on, and a covariance matrix according to 
\begin{equation}
\V{P}^{(0)}_i = \text{diag}([(\V{\sigma}_p^2)\transp,\Delta T^2 (\V{\sigma}_{a_{\text{init}}}^2)\transp,(\V{\sigma}_{a_{\text{init}}}^2)\transp])
\label{eq:prior_cov}
\end{equation}

\begin{figure*}[t]
\centering
\subfloat[ $r_\text{max}$ = 18 m]
{\includegraphics[scale=1]{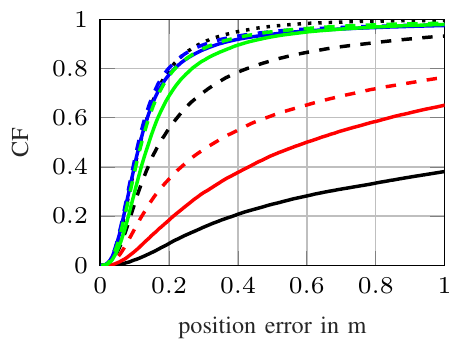}\label{fig:dyn_toy_a}}
\subfloat[$r_\text{max}$ = 18 m]{\hspace{-2mm}
{\includegraphics[scale=1]{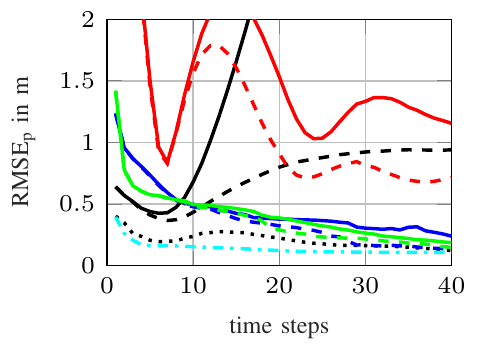}\label{fig:dyn_toy_b}}}
\subfloat[$r_\text{max}$ = 18 m]{\hspace{-2mm}
{\includegraphics[scale=1]{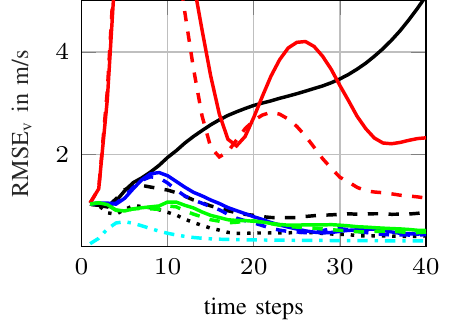}\label{fig:dyn_toy_c}}}
\subfloat[$r_\text{max}$ = 18 m]{\hspace{-2mm}
{\includegraphics[scale=1]{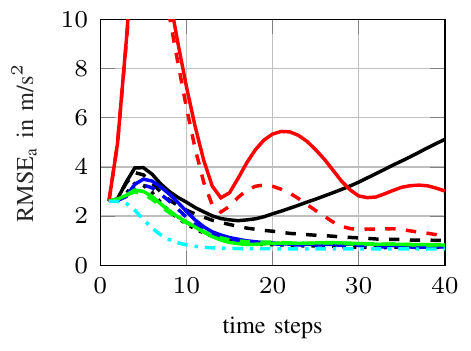}\label{fig:dyn_toy_d}}} \\
\subfloat[$r_\text{max}$ = $\infty$]
{\includegraphics[scale=1]{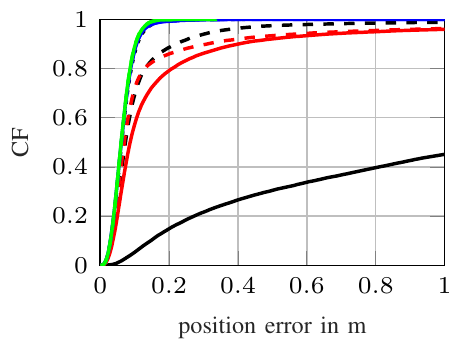}\label{fig:dyn_toy_e}}
\subfloat[$r_\text{max}$ = $\infty$]{\hspace{-2mm}
{\includegraphics[scale=1]{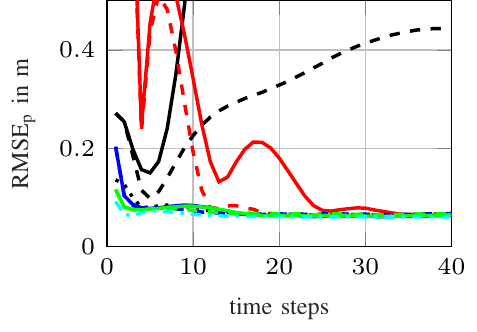}\label{fig:dyn_toy_f}}}
\subfloat[$r_\text{max}$ = $\infty$]{\hspace{-2mm}
{\includegraphics[scale=1]{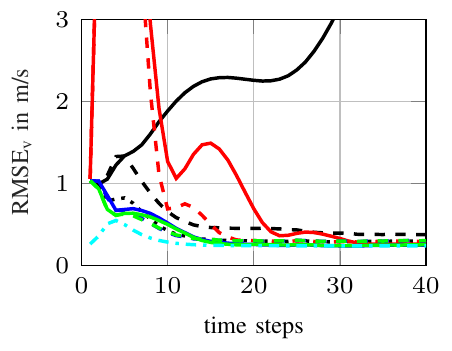}\label{fig:dyn_toy_g}}} 
\subfloat[$r_\text{max}$ = $\infty$]{\hspace{-2mm}
{\includegraphics[scale=1]{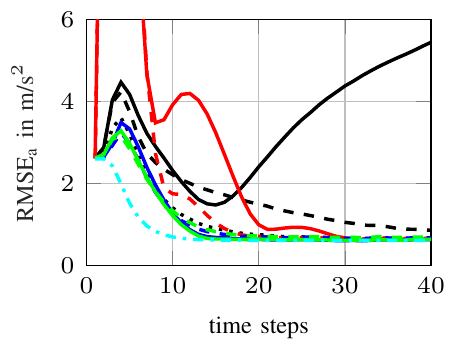}\label{fig:dyn_toy_h}}}\\
\includegraphics[scale=1]{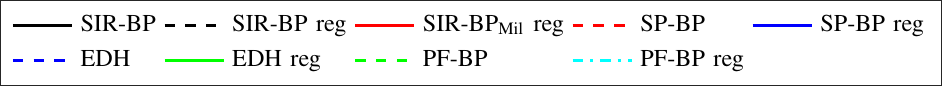}\hfill\\
\caption{Influence of the communication range $r_\text{max}$ on the performance in terms of accuracy for 5 agents and $\sigma = 0.1$~m for 200 simulation runs. We show the CF of the position error over the whole trajectory as well as the RMSE of the agent states at each time step, where we look separately at the position, velocity, and acceleration. The theoretical performance limit is given in terms of the PCRLB. Regularization is indicated by reg.}
\label{fig:dyn_4agents}
\end{figure*}

where  $\V{\sigma}_p^2 = [\sigma_{px}^2,\sigma_{py}^2,\sigma_{pz}^2]\transp$. We define the prior standard deviation of the position to be identical in all dimensions and set it to 20~m. For $\V{\sigma}_{a_{\text{init}}}^2$, we also define it to be identical in all dimensions. It is given as  $\V{\sigma}_{a_{\text{init}}}^2 = [(10\sigma_a)^2,(10\sigma_a)^2,(10\sigma_a)^2]\transp$.
The mean values $\overline{\V{v}}^{(0)}_i$ and $\overline{\V{a}}^{(0)}_i$, corresponding to velocity and acceleration respectively, are drawn from the zero-mean Gaussian distribution defined by the covariance matrix in \eqref{eq:prior_cov}. The mean value $\overline{\V{p}}^{(0)}_i$, corresponding to the position, is drawn uniformly in the support volume. For the SIR-BP algorithm, the particles representing the position are drawn uniformly in the support volume. In contrast, for the EDH filter and the PF-BP algorithm, the particles are drawn from the Gaussian prior distribution. One realization of the dynamic scenario with 20 agents and a communication range of $r_\text{max}$~=~18~m is given in Figure~\ref{fig:dyn_scen}. This figure also shows the anchors' placement at the corners of the support volume and the placement of a single anchor in the center. In addition, we indicate in color how many anchor measurements an agent has at each point of its trajectory. The setup is chosen such that each agent lies within the communication range of at least one anchor at each time step. For an agent to be fully localizable based on anchor measurements, one needs measurements from four different anchors where the positions of the anchors do not lay on a plane. As we see in Figure~\ref{fig:dyn_scen}, agents would not be localizable without cooperative measurements for most of the trajectories.

\begin{figure}[h]
\centering
\includegraphics[scale=1]{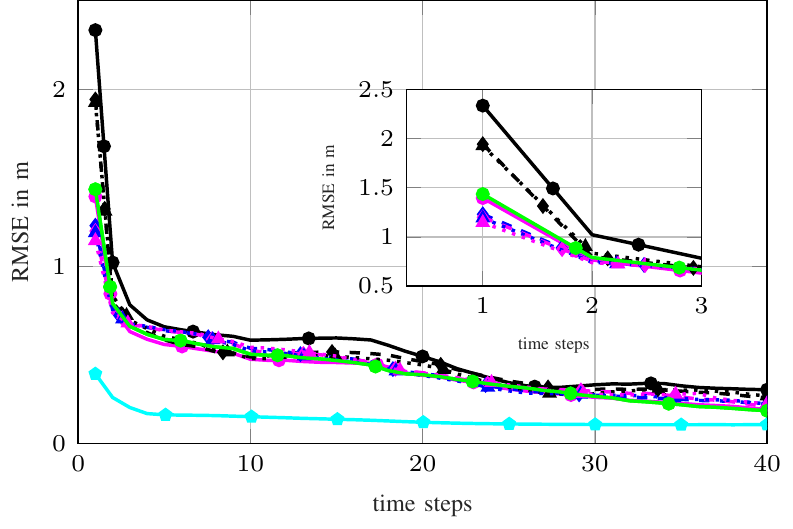}\\
\includegraphics[scale=1]{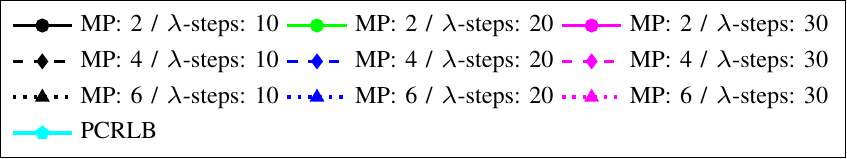}\hfill\\
\caption{Convergence behaviour of PF-BP with respect to message passing iterations and pseudo-time-steps. The results are averaged over 200 simulation runs. The setting corresponding to the green line is used for all other simulations.}
\label{fig:convPFBP}
\end{figure}

\begin{figure*}[t]
\centering
\subfloat[$k=1$, $r_\text{max}$ = 18 m]{
{\includegraphics[scale=1]{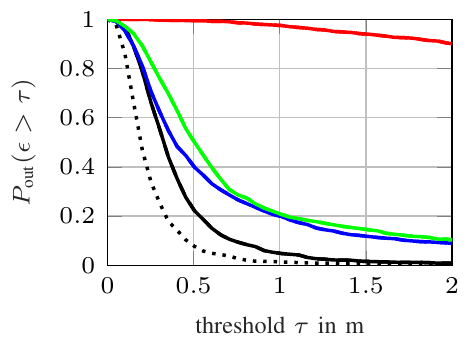}\label{fig:Pout_small_a}}}
\subfloat[$k=20$, $r_\text{max}$ = 18 m]{
{\includegraphics[scale=1]{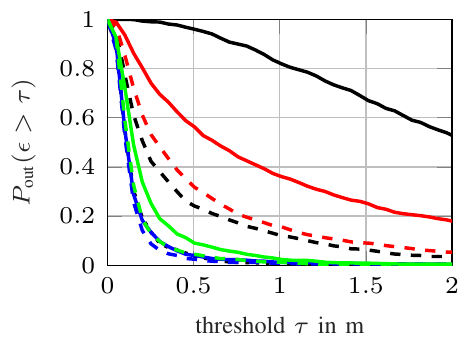}\label{fig:Pout_small_b}}}
\subfloat[$k=40$, $r_\text{max}$ = 18 m]{
{\includegraphics[scale=1]{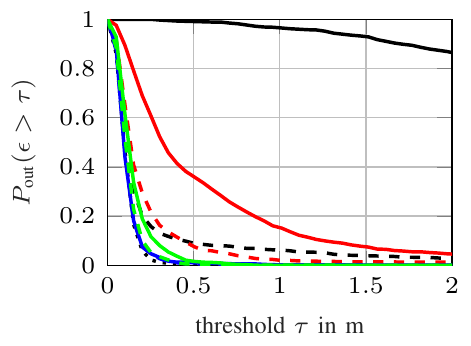}\label{fig:Pout_small_c}}} \\
\subfloat[$k=1$, $r_\text{max}$ = $\infty$]{
{\includegraphics[scale=1]{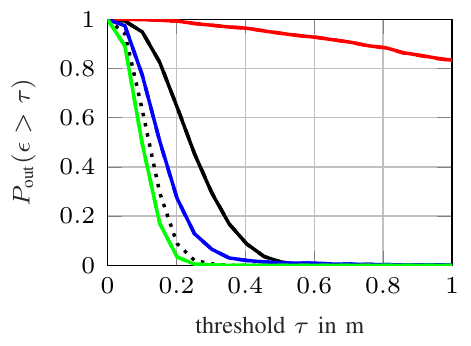}\label{fig:Pout_small_d}}}
\subfloat[$k=20$, $r_\text{max}$ = $\infty$]{
{\includegraphics[scale=1]{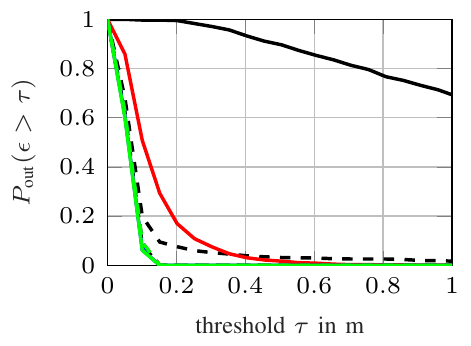}\label{fig:Pout_small_e}}}
\subfloat[$k=40$, $r_\text{max}$ = $\infty$]{
{\includegraphics[scale=1]{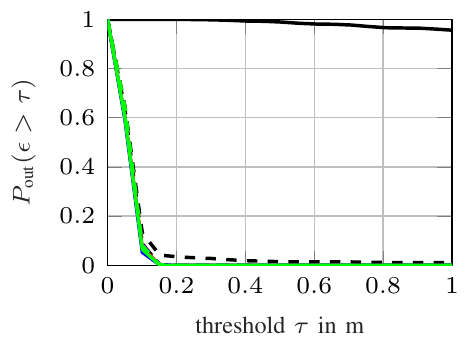}\label{fig:Pout_small_f}}}\\
\includegraphics[scale=1]{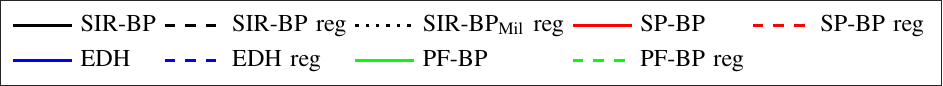}\hfill\\
\caption{Probability of outage of the position error for the investigated algorithms for the scenario with five agents. The first row shows the probability of an outage for a communication range of $r_\text{max}$=18~m, whereas the second row presents the probability of an outage for the fully connected case. It is evaluated at certain time-steps $k$. Regularization is indicated by reg.}
\label{fig:Pout_5agents}
\end{figure*}

\begin{figure*}[t]
\centering
\subfloat[CF of the overall trajectory]{\hspace{-2mm}
{\includegraphics[scale=1]{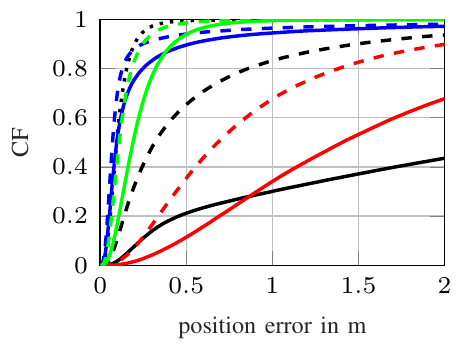}\label{fig:dyn_large_a}}}
\subfloat[Position RMSE]{\hspace{-2mm}
{\includegraphics[scale=1]{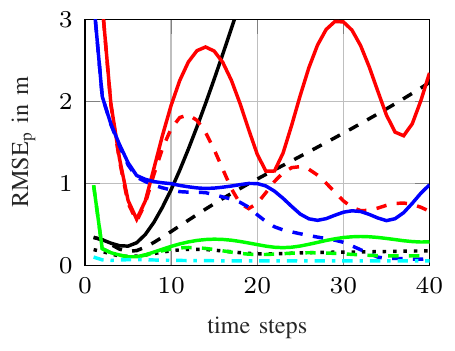}\label{fig:dyn_large_b}}}
\subfloat[Velocity RMSE]{\hspace{-2mm}
{\includegraphics[scale=1]{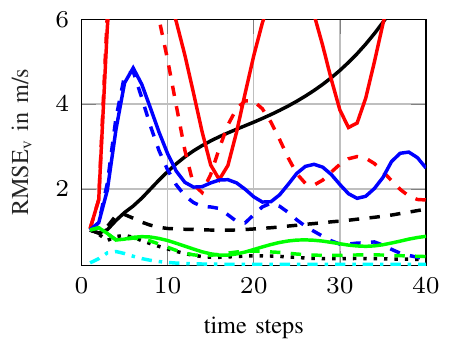}\label{fig:dyn_large_c}}}
\subfloat[Acceleration RMSE]{\hspace{-2mm}
{\includegraphics[scale=1]{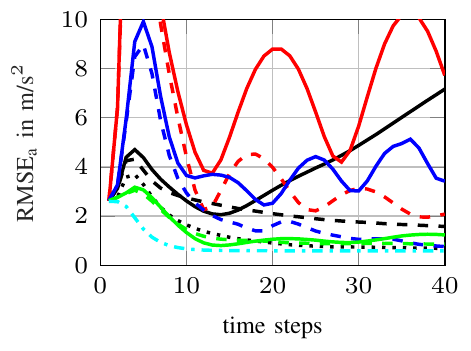}\label{fig:dyn_large_d}}} \\
\subfloat[$k=1$]{
{\includegraphics[scale=1]{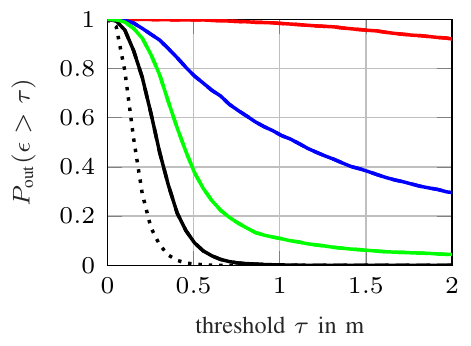}\label{fig:Pout_large_a}}}
\subfloat[$k=20$]{
{\includegraphics[scale=1]{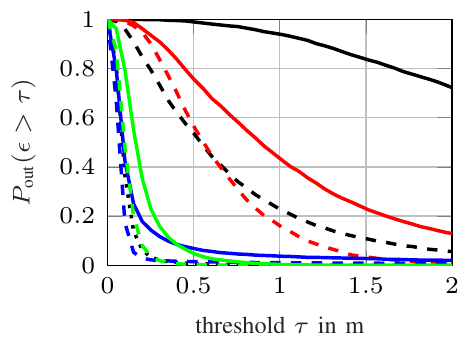}\label{fig:Pout_large_b}}}
\subfloat[$k=40$]{
{\includegraphics[scale=1]{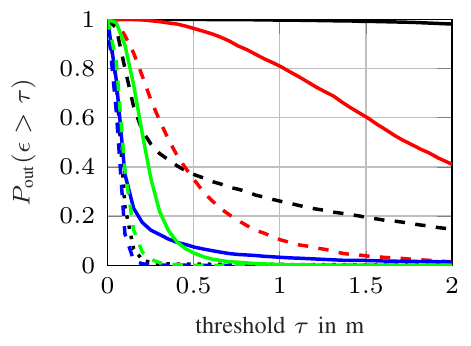}\label{fig:Pout_large_c}}}\\
\includegraphics[scale=1]{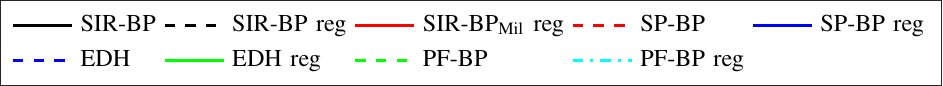}\hfill\\
\caption{Visualization of the performance of the investigated algorithms in terms of accuracy for the scenario with 20 agents and $r_\text{max}=18$~m. The first row shows the CF of the position error over the whole trajectory as well as the RMSE of the agent states at each time step, where we look separately at the position, velocity, and acceleration. The second row depicts the probability of outage of the position error at certain time-steps $k$. The results are given for $\sigma = 0.1$~m for 200 simulation runs. Regularization is indicated by reg.}
\label{fig:dyn_20agents}
\end{figure*}

We simulate 200 trajectories of the agents for $K=40$ time-steps. We use 20 $\lambda$-steps and 200 particles for the PF-based algorithms and 100\,000 particles for the SIR-BP algorithm. As an additional benchmark, we use 1\,000\,000 particles for the SIR-BP algorithm indicated as SIR-BP$_{\text{Mil}}$. We fix the number of MP-iterations to 2. More iterations would be more time-consuming, and the benefit regarding the convergence behavior of the BP-based algorithms would be negligible. Further insights regarding this topic is provided later on in this section.
Since it is common to use regularization to avoid particle degeneracy \cite{MeyerJSAC2015}, we investigate the impact of regularization on all presented methods. For that purpose, we regularize velocity and acceleration with $\sigma_{r_\text{vel}} = 0.15$~m/s and $\sigma_{r_\text{acc}}= 0.15$~m/s$^2$ for all investigate algorithms. This is done as follows: We define a Gaussian kernel with a covariance matrix 
\begin{equation}
\V{\Sigma}_r = \text{diag}([0,0,0,\sigma_{r_\text{vel}}^2,\sigma_{r_\text{vel}}^2,\sigma_{r_\text{vel}}^2,\sigma_{r_\text{acc}}^2,\sigma_{r_\text{acc}}^2,\sigma_{r_\text{acc}}^2]).
\end{equation}
For the UKF update and SP-BP, we add this covariance to the estimated covariance of each marginal state. Using for example \eqref{eq:PFBP_update}, it would result in
\begin{equation}
\V{P}_{i}^{(k)[u]}  = \V{P}_{i}^{(k|k-1)}  - \tilde{\V{K}}^{[u]} \tilde{\V{P}}_{zz} \tilde{\V{K}}^{[u] \text{T}} + \V{\Sigma}_r.
\label{eq:PFBP_reg}
\end{equation}
For the particle-based methods, we draw for each particle after resampling $\V{x}^{(k),m}_i$ a new particle $\acute{\V{x}}_i^{(k),m}$, which is distributed according to a Gaussian distribution with mean value $\V{x}^{(k),m}_i$ and covariance $\V{\Sigma}_r$. Results with regularization are indicated with dashed or dotted lines in the following figures and with ``reg'' in the legends.

\begin{table}[t]
  \caption{Runtime per time step for the results with 5 agents with respect to a joint and a distributed (distr.) processing. For the distributed processing, the results are given in runtime per agent.}
\begin{center}
  \begin{tabular}{l | c | r | r | r | r | r } 
& $r_\text{max}$ & SP-BP  &   EDH  & PF-BP   &  SIR-BP  &  SIR-BP$_{\text{Mil}}$  \\ \hline \hline
\multirow{2}{*}{joint} & 18	 &  3~ms &  10~ms &  50~ms  &  0.44~s   &   4.4~s                  \\\cline{2-7} 
& $\infty$   &  4~ms &  20~ms &  60~ms  &  0.51~s &   5.1~s         \\ \hline
\multirow{2}{*}{distr.} & 18 &  0.6~ms &  - &  10~ms  &  0.09~s  &   0.9~s \\\cline{2-7} 
& $\infty$   &  0.8~ms &  - &  12~ms  &  0.10~s  &   1.2~s                 \\ \hline
 \end{tabular}
\end{center}
\label{tb:dynamicTime}
\end{table}

\subsubsection{Scenario I}

We evaluate a scenario with 5 agents for different communication ranges $r_\text{max}$. For $r_\text{max}$~=~18~m, agents have at least one connection to an anchor, which is a similar scenario as given in Figure~\ref{fig:dyn_scen}. The results for that setting are given in Figure~\ref{fig:dyn_toy_a}-\ref{fig:dyn_toy_d} where we show the CF of the overall trajectory and the RMSE of position, velocity, and acceleration for each time step. We see clearly that the EDH filter and the proposed PF-BP algorithm outperform the SP-BP algorithm and the SIR-BP algorithm significantly in terms of accuracy without regularization. Table~\ref{tb:dynamicTime} shows the runtime per time-step for each algorithm with respect to a joint and a distributed processing. For a distributed processing, the runtime is given per time-step and agent. For a small number of agents and the chosen numbers of particles, the SP-BP algorithm outperforms all other methods in terms of runtime.

At the first few time-steps, some of the marginal posterior PDFs of the agent states are still multimodal, which can be well represented by the particles of the SIR-BP algorithm. Hence, the SIR-BP algorithm converges much faster to the ``correct mode'' of the posterior PDF leading to a much lower position error at the beginning of the agent trajectories (see Figure~\ref{fig:dyn_toy_b}). However, after a few steps, we can observe that the SIR-BP algorithm diverges in almost every simulation run since the chosen number of particles (100\,000) is still too small to sufficiently represent the 9-D agent state vectors. With regularization, the SIR-BP algorithm achieves a much better performance. However, we can still observe a significant bias in the RMSE, indicating that the chosen number of particles is still too low. With 1\,000\,000 particles and regularization, the SIR-BP algorithm reaches almost PCRLB level; however, with the cost of a significant increase of runtime (see Table~\ref{tb:dynamicTime}) making it not applicable for real-time applications and systems with memory restrictions. The small bias, that occurs, can be avoided using even more particles (not shown). The SP-BP algorithm also benefits from the regularization since it leads to faster convergence of the MMSE estimate over time towards the PCRLB. However, the achievable accuracy is still very low compared to the PCRLB. Furthermore, it was observed that the posterior covariance matrices provided by SP-BP are significantly overconfident (not shown). For both PF-based methods, regularization has only a slight impact. 

For a fully connected agent network (highly informative measurement models), we see clearly in Figure~\ref{fig:dyn_toy_e}-\ref{fig:dyn_toy_h} the superiority of both PF-based methods. The proposed PF-BP algorithm reaches the theoretical performance limit much faster compared to the other methods. The EDH filter reaches the PCRLBs after a few time-steps. The SP-BP algorithm needs significantly more time-steps until converging towards the PCRLBs. Using 100~000 particles, the SIR-BP algorithm obviously diverges with and without regularization in every simulation run. Even with 1~000~000 particles, the SIR-BP algorithm only converges if regularization is activated. Figure~\ref{fig:dyn_toy_f} shows that in this case, the SIR-BP algorithm also reaches the position PCRLB; however, due to the regularization, the velocity and acceleration RMSEs are biased. As a consequence of the large runtime and huge memory requirements, we do not present results with even more particles.

Both PF-based methods reach the PCRLBs without the need for regularization. Figure~\ref{fig:dyn_toy_g} shows that regularizing the PF-based methods only induces error biases to all states and is counterproductive for highly informative measurement models. Figure~\ref{fig:dyn_toy_g} also indicates that the SP-BP and SIR-BP algorithms benefit from the regularization since their estimates of velocity and acceleration need more time-steps to converge or even diverge without regularization. We conclude that regularization should be treated cautiously, as it has a sensitive effect on error biases. 

The runtimes of the investigated algorithms for both agent network are reported in Table~\ref{tb:dynamicTime}. They were determined based on a centralized and a distributed processing. The results indicate that even though PF-BP has a higher computation time compared to the EDH filter if processed centralized, the per-agent computations for a distributed processing are lower or of similar computation time.

In addition, we investigated the convergence behaviour of our proposed method with respect to $r_{\text{max}}=18$~m. Figure~\ref{fig:convPFBP} depicts the convergence over time-steps of the trajectory towards the PCRLB with regard to different MP iterations and different numbers of $\lambda$-steps. It can be observed that a larger number of $\lambda$-steps is always more beneficial than more MP iterations. Therefore, we fixed the number of MP iterations to 2 and the number of $\lambda$-steps to 20 for all simulations as mentioned in the beginning of this section. The result with this set of parameters is indicated in green.

Furthermore, we show in Figure~\ref{fig:Pout_5agents} the probability of outage ${P_{\text{out}}(\epsilon>\tau)}$ of the position error $\epsilon$, where $\tau$ is the position threshold in meters. We evaluate it at three time-steps $k \in \{1,20,40\}$. At $k=1$, we can see the benefits of the different algorithms. Figure~\ref{fig:Pout_small_a} shows for $r_{\text{max}}=18$~m at $k=1$, that the SIR-BP algorithm with 1\,000\,000 provides the most accurate results, followed by SIR-BP with 100\,000. This is because not every agent is localizable in the first step, and as mentioned above, SIR-BP can represent any PDF if enough particles are available. In Figure~\ref{fig:Pout_small_d}, there are no multimodalities in the position state due to the fully connected scenario. Therefore the unimodal approximation of the PF-BP algorithm is sufficient to represent the agent state correctly. Hence, it achieves higher accuracy than the SIR-BP with 1\,000\,000. For $k=20$, all particle-based methods have a similar performance except the SIR-BP algorithm without regularization. The estimates of SP-BP are still biased in Figure~\ref{fig:Pout_small_b}, whereas they are close to the optimum result in Figure~\ref{fig:Pout_small_e}. At the last step, we see that if converged, all algorithms perform approximately the same, which is equivalent to the results in Figure~\ref{fig:dyn_toy_f} where all investigated methods reach the PCRLB at the last time step.

\begin{table}[t]
  \caption{Runtime per time step for the results with 20 agents with respect to a joint and a distributed (distr.) processing. For the distributed processing, the results are given in runtime per agent.}
\begin{center}
  \begin{tabular}{ c | r | r | r | r | r } 
             & SP-BP &   EDH  & PF-BP   &  SIR-BP  &  SIR-BP$_{\text{Mil}}$ \\ \hline \hline
joint	    &  0.07~s   &  0.25~s & 0.9~s    &   3.6~s   &    40~s \\ \hline
distr.      &  0.004~s  &  -      &  0.05~s  & 0.18~s    &   2~s    \\ \hline
 \end{tabular}
\end{center}
\label{tb:dynamicTime20agents}
\vspace{-0.4cm}
\end{table}

\subsubsection{Scenario II}

In Figure~\ref{fig:dyn_20agents}, we show the results for 20 agents and a communication range of $r_\text{max}$~=~18~m. The results look similar to those given in Figure~\ref{fig:dyn_4agents} but with two major differences. 
At first, we can observe that none of the investigated methods reach the PCRLB with the defined parametrization. However, PF-BP has the smallest bias. Furthermore, we see that the estimates of the PF-based methods at $k=1$ differ significantly. Since the joint state now has 180 dimensions compared to the 45 dimensions of the scenario with five agents, the EDH filter has many more problems representing the state correctly. The PF-BP algorithm determines the marginal posterior PDFs of the agents and calculates the flow only based on a subset of the joint state, i.e., the state of agent $i$ and all other agents connected to it. Therefore the state dimension is much smaller, which also reduces the effect of particle degeneracy. This leads, with the same parameter setting, to a similar result to the one with five agents in Figure~\ref{fig:dyn_toy_b}. The discrepancy to the SIR-BP algorithm at $k=1$ shows that the PF-BP algorithm can not resolve multimodalities.  We can observe that all investigated methods benefit from the regularization for this scenario and the specific parameter setting. The RMSE of the PF-BP algorithm has a constant bias without regularization in Figure~\ref{fig:dyn_large_b}. This could be resolved with more particles, which increases the runtime. The same is true for the EDH filter. We can also see that the PF-based methods are the only ones that can reach the PCRLB within the time of the trajectory with a reasonable calculation time. 
The runtimes per time step are summarized in Table~\ref{tb:dynamicTime20agents} for a joint and a distributed processing. We see that the SIR-BP algorithm has a long runtime and is, therefore, unsuitable for real-time applications. The PF-BP algorithm also has a larger runtime than the EDH filter but only if processed jointly, hence making it suitable for real-time applications. SP-BP outperforms all other methods in terms of runtime but does not converge at all to the theoretical limit of the estimation performance.

Note that for highly informative prior distributions of the agent states at time $k=1$, the PF-based methods would still have higher accuracy than the SIR-BP and SP-BP algorithms. However, specifically for the SP-BP algorithm, the difference is significantly smaller.

In what follows, we summarize the advantages and disadvantages of the comparison methods and the proposed algorithm.
\begin{itemize}
\item The SIR-BP algorithm requires many particles to represent the posterior PDFs of the 9-D agent states correctly. Therefore, the algorithm has a long runtime and requires significant memory. However, the SIR-BP algorithm has the potential to correctly represent the posterior PDFs of the agent states asymptotically in the number of particles. It can therefore capture multimodalities in the posterior PDFs. 
\item The SP-BP algorithm has low computational demand and, therefore, a low run time. However, it shows slow convergence toward smaller RMSEs for high dimensional agent states over time. 
\item The particle-based EDH filter is suitable for small agent networks since it provides PCRLB-level position accuracy and has a low runtime. However, for larger networks, the convergence of the MMSE estimates over time is relatively slow, i.e., it needs many time-steps to reach PCRLB-level. Due to the joint state representation, it also does not scale well in the number of agents.
\item The proposed PF-BP algorithm provides high position accuracy at the PCRLB level and exhibits low running time per time step for distributed processing. It also converges quickly over time and scales well in the number of agents due to the possibility of a distributed implementation.
\end{itemize}

Regarding the communication overhead, we can draw the following conclusions: SP-BP and PF-BP use a Gaussian approximation, which means that Gaussian distributions represent the agent states. Therefore, each agent has to transmit only the mean value and the covariance corresponding to its belief instead of all particles, as is the case for SIR-BP. For PF-BP, each agent has to sample locally from that Gaussian distribution to perform the particle flow process in the measurement update step. The EDH cannot be implemented in a distributed manner, leading to the case where a central computation unit has to collect all measurements and perform the computation.

To make the advantages of the proposed method even clearer, the runtimes of the investigated algorithms were determined for centralized and distributed processing. The results indicate that even though PF-BP has a higher computation time compared to the EDH filter if processed centralized, the per-agent computations for a distributed processing are lower or of similar computation time.
%%%%%%%%%%%%%%%%%%%%%%%%%%%%%%%%%%%%%%%%%%%%
\section{Conclusion}
\label{sec:Conclusion}
%%%%%%%%%%%%%%%%%%%%%%%%%%%%%%%%%%%%%%%%%%%%%

We have proposed a Bayesian method based on belief propagation (BP) and particle flow for cooperative localization and navigation. Our method is particularly suitable for scenarios with high-dimensional agent states and informative nonlinear measurement models. To avoid particle degeneracy in such scenarios, invertible PF is used to compute BP messages. As a result, the proposed PF-BP algorithm can reach position accuracy at PCRLB level in a cooperative localization scenario with 9-D agent states and range measurements. Our numerical results demonstrate a reduced computational demand and memory requirement compared to the conventional SIR-BP algorithm and a particle-based EDH filter applied to cooperative localization. In addition, the communication overhead is reduced significantly with respect to SIR-BP and is comparable to SP-BP, which relies on a similar Gaussian representation. We performed simulations with different numbers of agents and communication ranges, demonstrating the superior estimation performance of the proposed PF-BP approach compared to state-of-the-art reference methods. We highlight the benefits and disadvantages of each investigated method in various scenarios. 

Possible future work is to extent the measurement model beyond Gaussian noise, like missed detections, clutter/false alarm measurements, and data association uncertainty of measurements \cite{VenLeiTerWit:RadarCon2021, Gaglione2022FusionMOT, BraGagSolRicGabLepNicWilBraWin:JSP2022, meyer2020_scalabelDA}, or to cooperative radio signal-based SLAM algorithm with highly informative measurement models \cite{LeiGreWit:ICC2019, LeiMey:Asilomar2020_DataFusion, KimTWC2020}.

\begin{appendices}

\section{Derivation of the PF Equation}
\label{app:A}

The drift term $\V{\zeta}(\V{x}^{(k)},\lambda)$ can be determined using the FPE, which is given as
\begin{align}
\frac{\partial f(\V{x}^{(k)};\lambda)}{\partial \lambda} = & -\nabla_{\V{x}}\transp (f(\V{x}^{(k)};\lambda) \V{\zeta}(\V{x}^{(k)},\lambda)) \nonumber \\
& + \frac{1}{2}\nabla_{\V{x}}\transp(f(\V{x}^{(k)};\lambda)\V{Q}(\V{x}^{(k)},\lambda))\nabla_{\V{x}}
\label{eq:Fokker-Planck-nonzero}
\end{align} 
where $\V{Q}(\V{x}^{(k)},\lambda)$ corresponds to the diffusion term. The solutions of \eqref{eq:Fokker-Planck-nonzero} for $\V{\zeta}(\V{x}^{(k)},\lambda)$ can be categorized into zero-diffusion, i.e., $\V{Q}(\V{x}^{(k)},\lambda) = 0$ \cite{daum2010exact,li2017particle} and nonzero-diffusion \cite{daum2013nonzeroDiffusion,daum2018Gromov}. The following two useful relations are used in the further derivation of the method:
\begin{enumerate}[leftmargin=*]
\item Using the chain rule of the divergence, the fist term in \eqref{eq:Fokker-Planck-nonzero} can be rewritten as
	\begin{align}
	&\hspace*{-2.5mm}\nabla_{\V{x}}\transp (f(\V{x}^{(k)};\lambda) \V{\zeta}(\V{x}^{(k)},\lambda)) \nonumber \\ 
	&\hspace*{-2mm}=\rmv f(\V{x}^{(k)};\lambda) \nabla_{\V{x}}\transp \V{\zeta}(\V{x}^{(k)}\rmv,\lambda) \rmv+\rmv (\nabla_{\V{x}}\transp  f(\V{x}^{(k)};\lambda))\V{\zeta}(\V{x}^{(k)}\rmv,\lambda).
	\label{eq:div_expand}
	\end{align} 
\item Using \eqref{eq:log-homotopy}, the left side of the FPE, namely the partial derivative with respect to $\lambda$, can be rewritten as
	\begin{align}
	&\frac{\partial f(\V{x}^{(k)};\lambda)}{\partial \lambda} \nonumber \\
	&\hspace*{3mm}= f(\V{x}^{(k)}|\V{x}^{(k-1)}) \left[\frac{\partial f(\V{z}^{(k)}|\V{x}^{(k)})^{\lambda}}{\partial \lambda}  \right]Z(\lambda)^{-1} \nonumber \\
	&\hspace*{5mm}+ f(\V{x}^{(k)}|\V{x}^{(k-1)}) \ f(\V{z}^{(k)}|\V{x}^{(k)})^{\lambda} \left[\frac{\partial Z(\lambda)^{-1} }{\partial \lambda}\right] \nonumber \\
	&\hspace*{3mm}= f(\V{x}^{(k)}|\V{x}^{(k-1)}) \ f(\V{z}^{(k)}|\V{x}^{(k)})^{\lambda} \nonumber \\ 
	&\hspace*{5mm}\times \text{log} f(\V{z}^{(k)}|\V{x}^{(k)}) \ Z(\lambda)^{-1} - f(\V{x}^{(k)}|\V{x}^{(k-1)}) \nonumber \\
	&\hspace*{5mm}\times f(\V{z}^{(k)}|\V{x}^{(k)})^{\lambda} Z(\lambda)^{-2} \left[\frac{\partial Z(\lambda) }{\partial \lambda}\right] \nonumber \\
	&\hspace*{3mm}= f(\V{x}^{(k)};\lambda) \left[ \text{log} f(\V{z}^{(k)}|\V{x}^{(k)}) - Z(\lambda)^{-1} \frac{\partial Z(\lambda) }{\partial \lambda}\right]\nonumber \\
	&\hspace*{3mm}= f(\V{x}^{(k)};\lambda) \left[ \text{log} f(\V{z}^{(k)}|\V{x}^{(k)}) -  \frac{\partial \text{log} Z(\lambda) }{\partial \lambda}\right].
	\label{eq:d_p}
	\end{align}
\end{enumerate}

By assuming zero-diffusion, \eqref{eq:Fokker-Planck-nonzero} simplifies to
\begin{equation}
\frac{\partial f(\V{x}^{(k)};\lambda)}{\partial \lambda} = -\nabla_{\V{x}}\transp (f(\V{x}^{(k)};\lambda) \V{\zeta}(\V{x}^{(k)},\lambda)).
\label{eq:Fokker-Planck-zero}
\end{equation}
Neglecting the derivative of the evidence $Z(\lambda)$ with respect to $\lambda$ \cite{daum2010exact}, and substituting \eqref{eq:d_p} and \eqref{eq:div_expand} into \eqref{eq:Fokker-Planck-zero}, we get
\begin{align}
\text{log} f(\V{z}^{(k)}|\V{x}^{(k)}) =& - [f(\V{x}^{(k)};\lambda)^{-1} \nabla_{\V{x}}\transp f(\V{x}^{(k)};\lambda)] \V{\zeta}(\V{x}^{(k)},\lambda) \nonumber \\
& -  \nabla_{\V{x}}\transp \V{\zeta}(\V{x}^{(k)},\lambda)
\end{align}
resulting in
\begin{align}
\nabla_{\V{x}}\transp \V{\zeta}(\V{x}^{(k)},\lambda) = & -\text{log} f(\V{z}^{(k)}|\V{x}^{(k)}) \nonumber \\
& -  (\nabla_{\V{x}} \text{log} f(\V{x}^{(k)};\lambda))\transp \V{\zeta}(\V{x}^{(k)},\lambda).
\label{eq:ODE}
\end{align}

\section{Implementation of the EDH filter}
\label{app:B}
Given \eqref{eq:stochPDE} and \eqref{eq:EDH}, we will describe here the state representation, matrices and vectors for the implementation of the EDH. Regarding \eqref{eq:EDH}, $\V{A}^{(k)}_\lambda$ and $\V{c}^{(k)}_\lambda$ are given as
\begin{align}
\V{A}^{(k)}_\lambda  = & -\frac{1}{2} \V{P}^{(k|k-1)} \V{H}^{(k) \text{T}} \nonumber \\
& \times (\lambda \V{H}^{(k)}   \V{P}^{(k|k-1)}  \V{H}^{(k) \text{T}}  \rmv + \rmv \V{R}^{(k)})^{-1} \V{H}^{(k)}  \label{eq:A_1}\\
\V{c}^{(k)}_\lambda = & (\V{I}_{N_{\text{D}} |\Set{C}|} \rmv + \rmv 2\lambda \V{A}) \nonumber \\
& \times [(\V{I}_{N_{\text{D}} |\Set{C}|} \rmv + \rmv \lambda \V{A}) \V{P}^{(k|k-1)} \V{H}^{(k) \text{T}} (\V{R}^{(k)})^{-1}  \nonumber \\
& \times (\V{z}^{(k)} \rmv + \rmv \V{\nu}^{(k)}) \rmv + \rmv \V{A} \overline{\V{x}}^{(k)}_{\lambda = 0}] \label{eq:b}
\end{align}
where $\V{\nu}^{(k)} = h(\overline{\V{x}}^{(k)}_{\lambda}) - \V{H}^{(k)}\overline{\V{x}}^{(k)}_{\lambda}$ and $\V{H}^{(k)} = \frac{\partial h(\V{x})}{\partial \V{x}} \left|_{\V{x} = \overline{\V{x}}^{(k)}_{\lambda}} \right.$, $h(\V{x})$ represents a shorthand notation to indicate all measurement hypotheses for all connected agents and anchors, and, $\overline{\V{x}}^{(k)}_{\lambda}$ represents the mean value of the state at pseudo time $\lambda$ and time step $k$ \cite{li2017particle}. For $\lambda = 0$, $\overline{\V{x}}^{(k)}_{\lambda = 0}$ corresponds to the mean value of the proposal PDF. Due to the Gaussian assumption, the proposal PDF is fully described by the mean value $\overline{\V{x}}^{(k)}_{\lambda = 0} \triangleq \overline{\V{x}}^{(k|k-1)}$ and the covariance matrix $\V{P}^{(k|k-1)}$ of the predicted agent state $\V{x}^{(k|k-1)}$. The predicted mean and the predicted covariance matrix can either be determined by the set of particles, i.e., $\overline{\V{x}}_{\lambda = 0}^{(k)} = (1/M)\sum_{m=1}^M\V{x}_{\lambda = 0}^{(k),m}$ and $\V{P}^{(k|k-1)} = (1/M)\sum_{m=1}^M (\V{x}_{\lambda = 0}^{(k),m}-\overline{\V{x}}_{\lambda = 0}^{(k)})(\V{x}_{\lambda = 0}^{(k),m}-\overline{\V{x}}_{\lambda = 0}^{(k)})\transp$ or by means of the Kalman-filter prediction equation as it will be described later on in this section. 

The particle representation $\{1/M,\V{x}^{(k),m}_{\lambda_l}\}^{M}_{m=1}$ of the joint state at pseudo-time-step $\lambda_l$ with $l \in \{1,\dots,N_\lambda\}$, where $N_\lambda$ is the maximum number of pseudo-time-steps, as well as the mean value of the particle representation can now be determined as 
\begin{align}
\V{x}^{(k),m}_{\lambda_l} & = \V{x}^{(k),m}_{\lambda_{l-1}} + \V{\zeta}(\V{x}^{(k),m}_{\lambda_{l-1}},\lambda_{l}) \Delta_l
 \label{eq:int_flow} \\
 \overline{\V{x}}^{(k)}_{\lambda_l} & = \overline{\V{x}}^{(k)}_{\lambda_{l-1}} + \V{\zeta}(\overline{\V{x}}^{(k)}_{\lambda_{l-1}},\lambda_{l}) \Delta_l
 \label{eq:int_flow_mean} 
\end{align} 
with $\Delta_l = \lambda_l - \lambda_{l-1}$ being the step size of the flow process between two consecutive pseudo time steps. This corresponds to the solution of \eqref{eq:stochPDE}.

To evaluate the proposal distribution corresponding to the particles \eqref{eq:int_flow} at the end of the flow ($\lambda = 1$), we make use of the invertible flow principle introduced in \cite{li2017particle}. Following that principle, the weights of the particles are recalculated based on the particle representation at the end ($\lambda = 1$) and the beginning ($\lambda = 0$) of the flow, i.e.,
\begin{equation}
w^{(k),m} \propto \frac{f(\V{x}_{\lambda = 1}^{(k),m}|\V{x}_{\lambda = 0}^{(k),m}) \ f(\V{z}^{(k)}|\V{x}_{\lambda = 1}^{(k),m})}{f(\V{x}_{\lambda = 0}^{(k),m})}.
\label{eq:jointWeights}
\end{equation}
Here, $\V{x}_{\lambda = 0}^{(k),m}$ is a particle sampled from the proposal PDF, represented by a Gaussian distribution. The posterior PDF of the joint agent state $\RV{x}^{(k)}$ is then represented by the set of weighted particles $\{ w^{(k),m},\V{x}_{\lambda = 1}^{(k),m}\}^{M}_{m=1}$. As final operation, we perform systematic resampling of the joint state resulting in the posterior PDF of the joint agent state at time $k$ given by $\{ 1/M,\V{x}^{(k),m}\}^{M}_{m=1}$ \cite{arulampalam2002tutorial} where we drop the index $\lambda$.

Similar to \cite{li2017particle} we calculate the posterior covariance matrix $\V{P}^{(k)}$ based on an unscented-Kalman-filter (UKF) update step \cite{merwe2000uncented, meyer2013sigma} at the sample-based mean value of the particle representation of the posterior PDF $\overline{\V{x}}_{\lambda = 1}^{(k)} = (1/M)\sum_{m=1}^M\V{x}_{\lambda = 1}^{(k),m}$ and the predicted covariance $\V{P}^{(k|k-1)}$. %Note that in \cite{ZhangMeyer2021MOT_PF}, the covariance matrix $\V{P}^{(k|k-1)}$ is calculated using the particle representation of the predicted state.
The predicted covariance matrix is given by
\begin{equation}
\V{P}^{(k|k-1)} = \tilde{\V{F}}\V{P}^{(k-1)} \tilde{\V{F}}\transp + \V{W}
\label{eq:EDH_prediction}
\end{equation}
where
\begin{align}
\tilde{\V{F}} & = \V{I}_{|\Set{C}|} \otimes \V{F} \label{eq:F_tilde} \\
\V{W} &= \V{I}_{|\Set{C}|} \otimes \V{Q} \label{eq:W} \\
\V{Q} &= \V{G} (\V{I}_{3} \odot \sigma_a^2) \V{G}\transp \label{eq:Q}
\end{align}
The update step is given as 
\begin{equation}
\V{P}^{(k)}  = \V{P}^{(k|k-1)} - \V{K}\V{P}_{zz}\V{K}\transp
\label{eq:EDH_update}
\end{equation}
with $\V{K}$ being the Kalman gain defined in \cite{merwe2000uncented,meyer2013sigma} and the measurement covariance matrix $\V{P}_{zz}$. More details on the UKF filter can be found in \cite{merwe2000uncented,meyer2013sigma}. 

It is possible to reduce the computation time of the EDH filter by comparing the rank of $\V{R}^{(k)}$ and $\V{P}^{(k|k-1)}$. If the rank of $\V{R}^{(k)}$ is larger than the rank of $\V{P}^{(k|k-1)}$, \eqref{eq:A_1} is reformulated using the Woodbury matrix identity.
\end{appendices}
\bibliographystyle{IEEEtran}
\renewcommand{\baselinestretch}{0.94}\small\normalsize
\bibliography{IEEEabrv,Main}

\end{document}